\documentclass[a4paper,11pt]{article}
\usepackage{cite}
\usepackage{amsmath,amssymb,graphicx,color,mathrsfs,subfigure,multicol}

\def\circa#1{\,\raise.3ex\hbox{$#1$\kern-.75em\lower1ex\hbox{$\sim$}}\,}
\makeatletter

\def\art{\@ifnextchar[{\eart}{\oart}}
\def\eart[#1]#2#3#4#5#6{{\rm #2}, {\em #3  #4} {\rm (#6) #5} ({\em #1})}
\def\hepart[#1]#2{{\rm #2, \em#1}}
\newcommand{\oart}[5]{{\rm #1}, {\em #2  #3} {\rm (#5) #4}}

\newcounter{alphaequation}[equation]
\def\thealphaequation{\theequation\hbox to
0.6em{\hfil\alph{alphaequation}\hfil}}
\def\eqnsystem#1{
\def\@eqnnum{{\rm (\thealphaequation)}}
\def\@@eqncr{\let\@tempa\relax \ifcase\@eqcnt \def\@tempa{& & &} \or
  \def\@tempa{& &}\or \def\@tempa{&}\fi\@tempa
  \if@eqnsw\@eqnnum\refstepcounter{alphaequation}\fi
\global\@eqnswtrue\global\@eqcnt=0\cr}
\refstepcounter{equation} \let\@currentlabel\theequation \def\@tempb{#1}
\ifx\@tempb\empty\else\label{#1}\fi
\refstepcounter{alphaequation}
\let\@currentlabel\thealphaequation
\global\@eqnswtrue\global\@eqcnt=0 \tabskip\@centering\let\\=\@eqncr
$$\halign to \displaywidth\bgroup \@eqnsel\hskip\@centering
$\displaystyle\tabskip\z@{##}$&\global\@eqcnt\@ne
\hskip2\arraycolsep\hfil${##}$\hfil& \global\@eqcnt\tw@\hskip2\arraycolsep
$\displaystyle\tabskip\z@{##}$\hfil
\tabskip\@centering&\llap{##}\tabskip\z@\cr}
\def\endeqnsystem{\@@eqncr\egroup$$\global\@ignoretrue} \makeatother

\def\endignore{}
\def\ignore #1\endignore{} 

\def\be{\begin{equation}}
\def\ee{\end{equation}}
\def\bea{\begin{eqnarray}}
\def\eea{\end{eqnarray}}

\def\endignore{}
\def\ignore #1\endignore{} 

\def\bd{\begin{displaymath}}
\def\ed{\end{diplaymath}}

\def\({\left(}
\def\){\right)}

\def\bea{\begin{eqnarray}}
\def\eea{\end{eqnarray}}
\def\be{\begin{equation}}
\def\ee{\end{equation}}

\def\bd{\boxdot}

\def\fnl{f_{\rm NL}}

\newcommand{\corr}[2]{\ensuremath{\langle \, #1 \; #2 \, \rangle }}

\newcommand{\corrbase}[1]{\ensuremath{\langle \, #1 \, \rangle }}

\newcommand{\corrk}[3]{\ensuremath{\langle \, #1_{\vec{#2}} \; #1_{\vec{#3}} \, \rangle }}

\newcommand{\kmes}[1]{\ensuremath{\frac{d^3 #1}{(2\pi)^3} \, }}

\newcommand{\random}[1]{ \ensuremath{ a_{\vec{#1}} } }
\newcommand{\randtens}[2]{ \ensuremath{ b_{\vec{#1}}^{#2} } }

\newcommand{\None}{\ensuremath{N_\phi}}

\newcommand{\deltaThreeD}[1]{\ensuremath{\delta ^{(3)} ( #1 ) \, } }

\newcommand{\zbar}{ \ensuremath{\bar{\zeta}} }
\newcommand{\gbar}{ \ensuremath{\bar{\gamma}} }

\newcommand{\treepowze}{ \ensuremath{\mathcal{P}_\zeta^{(0)}} }

\newcommand{\treenpow}[1]{ \ensuremath{P_{#1}^{(0)} }}

\newcommand{\powze}{ \ensuremath{\mathcal{P}_\zeta} }

\newcommand{\npow}[1]{ \ensuremath{P_{#1} }}

\def\tr{{\rm tr}\,}

\newcommand{\kv}{ \ensuremath{\vec{k}} }
\newcommand{\pv}{ \ensuremath{\vec{p}} }
\newcommand{\qv}{ \ensuremath{\vec{q}} }
\newcommand{\xv}{ \ensuremath{\vec{x}} }
\newcommand{\yv}{ \ensuremath{\vec{y}} }
\newcommand{\zv}{ \ensuremath{\vec{z}} }

\begin{document}

\hfill 2 February 2011 \hspace{1.0cm}

\vspace{0.5cm}

\begin{center}
{\LARGE \bf
Inflationary Correlation Functions\\[1.0ex] without Infrared Divergences
}\\[1cm]

{
{\large\bf Mischa Gerstenlauer$^{\dagger}$\footnote{E-mail:  
m.gerstenlauer@thphys.uni-heidelberg.de}
\,,\,    
Arthur Hebecker$^{\dagger}$\footnote{E-mail:  
a.hebecker@thphys.uni-heidelberg.de}
\,,\,    
Gianmassimo Tasinato$^{\dagger\star}$\footnote{E-mail:  
gianmassimo.tasinato@port.ac.uk}
}
}
\\[7mm]
{\it $^{\dagger}$ Institut f\"ur Theoretische Physik, Universit{\"a}t 
Heidelberg,\\
 Philosophenweg  19,
69120 Heidelberg, Germany
}\\[1cm]
\vspace{-0.3cm}
{\it $\star$ Institute of Cosmology and Gravitation,
 University of Portsmouth,\\ Dennis Sciama Building,
Portsmouth, PO1 3FX, UK}

\vspace{1cm}

{\large\bf Abstract}

\end{center}
\begin{quote}
Inflationary correlation functions are potentially affected by infrared 
divergences. For example, the two-point correlator of curvature perturbation 
at momentum $k$ receives corrections $\sim\ln(kL)$, where $L$ is the size of 
the region in which the measurement is performed. We define infrared-safe 
correlation functions which have no sensitivity to the size $L$ of the box 
used for the observation. The conventional correlators with their familiar 
log-enhanced corrections (both from scalar and tensor long-wavelength modes) 
are easily recovered from our IR-safe correlation functions. Among other 
examples, we illustrate this by calculating the corrections to the 
non-Gaussianity parameter $f_{NL}$ coming from long-wavelength tensor modes. 
In our approach, the IR corrections automatically emerge in a resummed, 
all-orders form. For the scalar corrections, the resulting all-orders 
expression can be evaluated explicitly.
\end{quote}

\bibliographystyle{unsrt}

\section{Introduction}

Infrared divergences associated with the inflationary power spectrum are 
a long-standing issue \cite{Mukhanov:1982nu, Vilenkin:1982wt,
Linde:1982uu, Starobinsky:1982ee, Vilenkin:1983xp, Linde:1990xn}
which has more recently received a lot of attention following \cite{
Weinberg:2005vy, Weinberg:2006ac}. Our focus is on divergences 
which directly affect the power spectrum. According to \cite{Lyth:2007jh} 
such divergences are cut off by the size $L$ of the observed patch of the late universe.
We are going to develop and generalize the analyses of \cite{Giddings:2010nc,
Byrnes:2010yc} (see also \cite{senatore-talk}). This approach emphasizes 
the way in which long-wavelength perturbations do (or do not) influence 
locally measured inflationary spectra. It may be related to earlier 
proposals of \cite{Urakawa:2009gb,Urakawa:2009my} and is clearly in line 
with at least part of the subsequent discussion in \cite{Urakawa:2010it}.

Infrared divergences explicitly arise in loop corrections to inflationary 
observables (using e.g.\ the in-in formalism) or through the nonlinear 
dependence of curvature perturbation on fluctuations of an underlying scalar
field (e.g.\ in the $\delta N$ approach). The vast amount of literature on the
subject (see e.g.~\cite{Starobinsky:1986fx, Starobinsky:1994bd, 
Sasaki:1992ux, Suzuki:1992gi, Gong:2001he, Enqvist:2008kt, Seery:2008ms, 
Urakawa:2010kr, Seery:2009hs, Durrer:2009ii, Burgess:2009bs, 
Rajaraman:2010zx, Burgess:2010dd,Senatore:2009cf, Kahya:2010xh, Kuhnel:2010pp, 
Finelli:2010sh, Prokopec:2010be, Cogollo:2008bi, Riotto:2008mv, 
Polyakov:2007mm, vanderMeulen:2007ah, Sloth:2006az, Linde:2005ht, Gong:2010yk}) 
 has recently been reviewed in \cite{Seery:2010kh}.
More or less by definition, IR divergences are due to modes which have a 
much longer wavelength than the characteristic scale of the problem under 
consideration. Focussing on correlation functions, it is then clear that 
such IR modes left the horizon earlier than the modes which are directly 
accessed via the correlation function at a given scale. Hence a generic effect 
of these IR modes is a modification of the background in which other modes 
propagate and are eventually observed.

Much debate has been raised on how to interpret such IR divergences. At least
in the case of single-field slow-roll inflation, it is now widely accepted 
that log-divergent integrals over soft modes have to be cut off at a $k_{min} 
\sim 1/L$, where $L$ is the typical size of the `box' in which the observer 
measures some correlator at momentum $k\gg 1/L$ \cite{Lyth:2007jh} (see also 
\cite{Lyth:2005fi, Boubekeur:2005fj, Lyth:2006gd, Bartolo:2007ti}). It is 
thus more appropriate to talk about a log-enhancement rather than a 
log-divergence. The suggestion that long-wavelength modes can be absorbed 
in the background and hence do not affect the correlator at momentum $k$
has been put forward long ago \cite{Unruh:1998ic, Geshnizjani:2005rg, 
Geshnizjani:2002wp, Geshnizjani:2003cn, Kofman:1986wm, Linde:1994yf}. 
In this sense, the absence of IR divergences in situations where 
$L$ is not much larger than $k$ may have been apparent to many authors 
even before Lyth's paper \cite{Lyth:2007jh} of 2007.

We note that backreaction of long-wavelength modes in quasi-de Sitter space-times has been considered also in other contexts \cite{Mukhanov:1996ak, Abramo:1997hu, Unruh:1998ic, Geshnizjani:2002wp, Geshnizjani:2003cn, Geshnizjani:2005rg,Tsamis:1992sx, Miao:2010vs, Koivisto:2010pj, Myhrvold:1983hu, Ford:1984hs, Antoniadis:1985pj, Antoniadis:1986sb}, most notably in attempts to compensate the cosmological constant or to explain the current accelerated expansion of the Universe. We have nothing to say concerning these topics and refer the reader to \cite{Seery:2010kh} for a more extensive compilation of the relevant literature. 

\smallskip

In this paper, we analyse IR effects associated with the backreaction of long-wavelength scalar and tensor modes in inflationary backgrounds, in the spirit of \cite{Giddings:2010nc,Byrnes:2010yc}. This approach is related to the consistency relations \cite{Maldacena:2002vr, Creminelli:2004yq, Seery:2008ax}. Developing and generalizing a suggestion made in \cite{Byrnes:2010yc} (see also \cite{Urakawa:2010it}), we propose an IR-safe definition of correlation functions involving curvature fluctuations. In doing so, we remove any sensitivity to modes that have a much longer wavelength than the scale at which the correlator is probed. The essential idea is to make use of the proper invariant distance on the reheating surface, where the curvature perturbation is evaluated. Considering two points on this surface, the dependence of the (physical) invariant distance on the coordinate vector, corresponding to the separation of the two points, is affected by long-wavelength contributions from geometrical quantities, namely the curvature and tensor perturbations. The misidentification of the distance due to long-wavelength modes is precisely the origin of IR effects. Consequently, by using the proper invariant distance, it is possible to construct $n$-point functions for the curvature perturbation that are free from the effect of long-wavelength modes and, hence, free from IR divergences associated with these contributions.

We show how to relate $n$-point functions, calculated in terms of the invariant distance, to the conventionally defined $n$-point functions. This allows us to provide closed expressions for the latter that manifestly exhibit the dependence on long-wavelength modes. As a consequence, in our approach the IR corrections to $n$-point functions automatically emerge in a resummed, all-orders form. When expanded at leading order in terms of long-wavelength modes, we recover the familiar log-enhanced, IR sensitive contributions. We apply our approach to the analysis of the two- and three-point functions for the curvature perturbation in slow-roll, single field inflation. The leading IR corrections to the power spectrum appear as log-enhanced contributions, multiplied by the power spectrum and second order slow-roll parameters. Furthermore, our resummed, all-orders expression allows us to evaluate IR corrections in a non-perturbative way by using statistical properties of the integrated long-wavelength fluctuation. We apply this framework to specific inflationary set-ups, obtaining a complete expression that includes all contributions of scalar long-wavelength modes to the power spectrum. Regarding the bispectrum, we derive the complete expression for long-wavelength scalar and tensor contributions to $\fnl$. We then expand the result at leading order in slow-roll, showing that tensor modes dominate the slow-roll expansion and provide the leading log-enhanced contributions to non-Gaussianity. Contrary to the power spectrum, we find that the leading order correction to $\fnl$ is suppressed only by first order slow-roll parameters.

We also show that, in all cases where the $\delta N$-formalism is applicable, our results can be equivalently obtained in terms of a suitable generalization of the $\delta N$-formalism, extending the discussion of \cite{Byrnes:2010yc}. In the present work, we include the effects of graviton long-wavelength modes, and we explain how to calculate IR contributions to arbitrary $n$-point functions involving curvature perturbations.

Log-enhanced contributions to inflationary observables, both in the in-in formalism and in the $\delta N$-formalism, have received much attention over the past few years. In the case of the $\delta N$-formalism, they have been associated with infrared divergences of the so called $C$-loops, and have been calculated mostly in terms of a diagrammatic expansion \cite{Boubekeur:2005fj,Zaballa:2006pv,Seery:2007wf,Byrnes:2007tm}. Although our approach is related, it is conceptually and technically different. We derive IR contributions directly from geometrical quantities. These corrections appear automatically in a resummed, all-orders form and do not need any diagrammatic expansion. Following arguments given in \cite{Giddings:2010nc}, the presented derivation of IR effects from the geometry of the reheating surface matches IR contributions due to quantum loop effects of long-wavelength modes calculated \`a la Weinberg \cite{Weinberg:2005vy,Weinberg:2006ac}, although a direct comparison is beyond the scope of the present paper.

\smallskip

The paper is organized as follows: In sec.~\ref{sec:geometry_of_the_reheating_surface}, we present the definition of IR-safe correlation functions and we show the emergence of IR corrections in a resummed, all-orders form by relating conventional correlation functions to their IR-safe equivalents. Furthermore, we show for scalars how this expression can be evaluated explicitly. In sec.~\ref{sec:alternative_way}, we give an alternative approach in single field, slow-roll inflation in the language of the $\delta N$-formalism. In sec.~\ref{sec:power_spectrum} and sec.~\ref{sec:3_point_bispectrum}, we apply our results to the power spectrum and to the non-Gaussianity parameter $\fnl$, respectively. We draw our conclusions in sec.~\ref{sec:conclusions}.

\section{Geometry of the reheating surface}

\label{sec:geometry_of_the_reheating_surface}

In this section we provide a physical interpretation for the appearance  of log-enhanced correction to inflationary correlation functions, developing and generalizing \cite{Giddings:2010nc, Byrnes:2010yc}. We start from the assumption that the reheating surface (or any other surface of constant energy density after the end of inflation), viewed as a metric manifold, represents in principle a physical observable. Neglecting vector modes, the metric of this surface can be written as
\be
  \label{eq:geometry_metric_3_surface}
  ds^2_3 = e^{2\zeta(\xv)} \left( e^{\gamma(\xv)} \right)_{ij} dx^i dx^j \qquad .
\ee
We choose a gauge where the symmetric matrix $\gamma$ is traceless and $\partial_i \gamma_{ij}=0$. Of course, $\zeta$ is accessible only indirectly, e.g.\ via $\delta T/T$ of the CMB radiation, but for the present paper we simply assume that this does not limit its observability. Considering, for example, single field slow-roll inflation, the expressions for $\zeta$ and $\gamma_{ij}$ (in momentum space) read \cite{Riotto:2002yw} 
\begin{align}\label{inexzg}
 \zeta(\qv) &= \frac{\None (q) \, H(q)}{\sqrt{2 q^3}} \; \random{q}  &
 \gamma_{ij}(\qv) &= \sum\limits_{s=+,\times} \frac{H(q)}{\sqrt{q^3}} \; \epsilon^s_{ij}(\qv) \; \randtens{q}{s}  \quad .
\end{align}
These quantities are conserved on superhorizon scales \cite{Salopek:1990jq, Maldacena:2002vr, Lyth:2004gb}. In the equations above, $\random{q}$ and $\randtens{q}{s}$ are normalized Gaussian random variables and $s$ is the helicity index for gravitational waves. The polarization tensors $\epsilon^s_{ij}$ are chosen to satisfy the transversality and tracelessness conditions, as well as an orthogonality relation\footnote{We use conventions such that the Fourier transform reads $\zeta (\xv) = \int \kmes{k} \; e^{i\kv\xv} \zeta (\kv)$~. The Gaussian random variables $\random{k}$ and $\randtens{k}{s}$ have zero mean and variance $\corr{\random{k}}{\random{p}} = (2\pi)^3 \; \deltaThreeD{\kv + \pv}$ and $\corr{\randtens{k}{s}}{\randtens{p}{s'}} = (2\pi)^3 \; \deltaThreeD{\kv + \pv} \; \delta^{ss'}$~. The polarization tensor for gravitational waves satisfies $\epsilon^s_{ii} (\kv) = 0 = k_i \epsilon^s_{ij} (\kv)$ and the orthogonality relation $ \sum_{ij} \epsilon^{s}_{ij} (\kv) \; \epsilon^{s'}_{ij} (-\kv) = 2\delta_{ss'}$~.}. Furthermore, $\None (q) = V/(dV/d\phi)$ and $H(q)=\sqrt{V(\phi)/3}$ with both quantities evaluated at the time of horizon exit of the mode $q$.

Although, for definiteness, we focus on slow-roll inflation, the particular expressions for $\zeta$ and $\gamma_{ij}$  given above are not essential for the formalism presented in this section. Consequently, our arguments are largely independent of the specific inflationary set-up under consideration. Important consequences for the $n$-point functions can be derived just from the geometry of the reheating surface specified above. Focussing on corrections to the curvature perturbation $\zeta$, we start by discussing the power spectrum and then generalize to spectra of $n$-point functions, for arbitrary $n$.

\subsection{The power spectrum}

Using its definition, $\corrk{\zeta}{k}{p} = (2\pi)^3 \deltaThreeD{\kv+\pv} 2\pi^2  \, \powze (k) / k^3$, the power spectrum can be written as the Fourier transform of the correlation function in real space:
\be
 \label{eq:geometry_power_spectrum}
 \powze (k)  =  \frac{k^3}{2\pi^2} \int\limits d^3 y \; e^{-i\kv\yv} \, 
\corr{\zeta(\xv)}{\zeta(\xv + \yv)} \qquad .
\ee
Since we want to interpret this formula as a practical prescription for the measurement of the power spectrum, we do not view $\langle\cdots\rangle$ as an abstract ensemble average. Instead, the averaging is over pairs of points separated by a coordinate-vector $\vec{y}$ within a certain part of the reheating surface. In other words, we are averaging over the location $\vec{x}$ of such pairs. This prescription clearly relies on a certain parameterization of the reheating surface and is hence gauge dependent. Nevertheless, given the gauge choice made earlier, the resulting $\powze(k)$ is a well-defined observable.

An observer is not able to probe the whole inflationary region. We assume that the observable patch is a box of volume $L^3$ to which the $\vec{x}$-averaging is  restricted. While this may not be immediately apparent from \eqref{eq:geometry_power_spectrum}, the power spectrum measured by a given observer depends on the box-size $L$. Qualitatively, this can be seen as follows:

Focus on a certain momentum $k$. Due to the Fourier transform, the power spectrum at this $k$ is determined by the behavior of the $\corr{\zeta(\xv)}{\zeta(\xv + \yv)}$ as a function of $y$ in the region $y\sim 1/k$ (here $y=\sqrt{\delta_{ij}\, y^i y^j}$ is the length of $\yv$). However, at different $\xv$ the same value of $y$ may correspond to different physical (invariant) distances between points $\xv$ and $\xv+\yv$ at which $\zeta(\xv)$ and $\zeta(\xv + \yv)$ are evaluated. The reason for this is the long-wavelength background
\begin{align}
 \label{eq:geometry_definition_x_dep_background}
 \zbar (\xv) &= \int\limits_{L^{-1} < \, q\, \ll k} \kmes{q} \; e^{i\qv\xv} \; \zeta (\qv)  
 &  \gbar_{ij} (\xv) &= \int\limits_{L^{-1} < \, q\, \ll k} \kmes{q} \; e^{i\qv\xv} \; 
 \gamma_{ij} (\qv) \quad ,
\end{align}
which varies significantly as $\xv$ varies over a box of (sufficiently large) size $L$. Indeed, the physical distance between the points $\xv$ and $\xv+\yv$ appearing in the average is given by $z^2 = e^{2\zbar} \left( e^{\gbar} \right)_{ij} y^i y^j$. Moreover, this mismatch between $y$ and the true distance $z$ grows with $L$. This effect is at least one of the origins of the familiar IR-problems of inflationary correlation functions. At leading order, IR-problems originate precisely from this effect.

To be more precise, we somewhat jump ahead and note that $\langle
\bar{\zeta}^2\rangle\sim (N_\phi H)^2\ln(kL)$, with a similar formula 
holding for $\bar{\gamma}$. In other words, the expectation value of 
$\bar{\zeta}^2$ grows logarithmically with $L$ because of the summation 
over modes between $1/L$ and $k$ involved in its definition. Thus, the 
effect of these backgrounds on $\zeta$-correlators at the scale $k$ can 
become large if the logarithm overcomes the suppression by the tree-level
power spectrum $(N_\phi H)^2$. Such a potentially large effect can come 
only from the factors $e^{2\zbar}$ and $e^{\gbar}$ relating the coordinate 
distance $y$ and the invariant distance $z$, as explained above. If we 
are able to remove this effect from the definition of the power spectrum, 
then we have removed all IR effects at the {\it leading-logarithmic} order.
By this we mean all corrections involving as many powers of $\ln(kL)$ as 
of the suppression factor $(N_\phi H)^2$, at leading order in slow-roll. 

To avoid this (leading-logarithmic) $L$-dependence (or IR-sensitivity), we 
propose to use the \textit{invariant} distance $z$ for the definition of 
the curvature correlator \cite{Byrnes:2010yc,Urakawa:2010it}. The background 
contains, by its very definition, only modes much longer than the relevant 
scales $y \sim 1/k$. Hence the background is smooth at the scale $y$. Its 
presence corresponds to a (constant) coordinate transformation $\yv\to\zv$:
\be
 z^i = e^{\zbar} \left( e^{\gbar/2} \right)_{j}^i \, y^j  \qquad .
\ee
The invariant distance $z=\sqrt{\delta_{ij} \; z^i z^j}$ represents the 
\textit{physical} separation of the points $\xv$ and $\xv+\yv$, 
independently of the location $\xv$ and the background in its surroundings. 

Thus, the correlator $\corr{\zeta(\xv)}{\zeta(\xv+ e^{-\zbar (\xv)} \, 
e^{-\gbar (\xv)/2} \, \zv)}$ involves an average over pairs of points that are 
separated by a certain invariant distance $z$. The $z$-dependence of this 
correlator is then a background-independent object. To make this even more 
apparent, we spell out the exact prescription for obtaining this correlator: 
The basic step consists in picking a pair of points from the reheating 
surface which are separated by an invariant distance $z$ and multiplying the 
corresponding values of $\zeta$. This in itself is not background independent 
since the background can shift $\zeta$ by a constant. However, once we 
restrict our interest to the $z$-dependence of this product of $\zeta$-values, 
any such constant drops out. Hence, the $z$-dependence of $\corr{\zeta(\xv)}
{\zeta(\xv+ e^{-\zbar (\xv)} \, e^{-\gbar (\xv)/2} \, \zv)}$ is indeed an IR-safe 
quantity: While the average is in practice over a certain region of size 
$L$, the expectation value is independent of where we are in this 
region. It can therefore not depend on the size $L$ of the underlying region. 
To say it yet in another way: Single-field inflation ends in the same way 
in every part of the universe and hence a correlator, defined in a purely
local manner, can not depend on the size of the region from which the 
sample of pairs of points is chosen. 

Consequently, we can define an IR-safe power spectrum, that we denote 
$\treepowze$, by
\be
  \label{eq:geomety_tree_power_spectrum}
  \treepowze (k) = \frac{k^3}{2\pi^2} \int\limits d^3 z \; e^{-i\kv\zv} \, \left\langle \; \zeta(\xv) \; \zeta(\xv+ e^{-\zbar(\xv)} \, e^{-\gbar(\xv)/2} \, \zv)  \right\rangle  \qquad .
\ee
This Fourier transform at scale $k$ is only sensitive to the $z$-dependence 
of the correlator in the region $z\sim 1/k$. It is hence IR-safe by the 
arguments given above. 

The expression for the original IR-sensitive power spectrum $\powze$ given in \eqref{eq:geometry_power_spectrum} follows from comparing eq.~\eqref{eq:geometry_power_spectrum} and eq.~\eqref{eq:geomety_tree_power_spectrum}. Starting from eq.~\eqref{eq:geometry_power_spectrum}, we express the vector $\yv$ in terms of the vector $\vec{z}$. Notice that this also affects the argument of the exponential. Then, we perform a coordinate transformation $d^3 y \rightarrow d^3 z$ in order to bring the integral in a form similar to eq.~\eqref{eq:geomety_tree_power_spectrum}. As a final step, we can express the result in terms of the IR-safe power spectrum evaluated at $ e^{-\zbar (\xv)} e^{-\gbar (\xv)/2}  \kv$~. A detailed calculation can be found in Appendix \ref{app:calculation_IR_safe_power_spectrum}. The result reads
\be
  \label{eq:geometry_comparison_power_spectra}
  \powze (k) = \left\langle \; \left[  \left(e^{-\gbar (\xv)}\right)_{ij} \hat{k}_i\hat{k}_j \right]^{-3/2}  \; \treepowze \left( e^{-\zbar (\xv)} e^{-\gbar (\xv)/2}  \kv \,\right)  \; \right\rangle \qquad .
\ee
The vector $\hat{k}$ is a unit-vector in $\kv$-direction and the average is performed  over the background quantities $\zbar(\xv)$ and $\gbar_{ij}(\xv)$. Neglecting tensor fluctuations in the equation above, we recover our result \cite{Byrnes:2010yc} for corrections to the power spectrum due to scalar fluctuations, i.e.\ $\powze (k) = \corrbase{\treepowze ( k e^{-\zbar})} $. Let us point out the presence of a prefactor, containing only tensor fluctuations, in eq.~\eqref{eq:geometry_comparison_power_spectra}. It is originating from the coordinate transformation $d^3 y \rightarrow d^3 z$ in the comparison of the two spectra. While scalar fluctuations receive a contribution from this transformation, tensor fluctuations do not, due to the fact that $\det e^\gamma = 1$. Expanding to leading non-trivial order in the background yields
\be
 \label{eq:geomety_expansion_power_spectrum}
 \powze (k) =  \left( 1 - \frac{1}{20} \corrbase{{\rm tr} \,\gbar^2} \frac{d}{d\ln k}  + \frac{1}{2} \corrbase{\zbar^2} \frac{d^2}{d(\ln k)^2} \right) \treepowze (k) \qquad ,
\ee
in agreement with \cite{Giddings:2010nc} (see also Section \ref{sec:power_spectrum}). Here, we used the zero mean condition $\corrbase{\zbar(x)} = 0=\corrbase{\gbar_{ij} (\xv)}$, which can always be realized by a rescaling of coordinates. In principle, one may choose coordinates where this is not the case. But it is rather natural to assume that an observer would specify coordinates in such a way that his observable patch is not affected by a constant background shift. In the particular case of slow-roll inflation, both corrections in eq.~\eqref{eq:geomety_expansion_power_spectrum} are of the same order. While, according to the scalar-to-tensor ratio, ${\rm tr}\,\gbar^2$ is more slow-roll suppressed than $\zbar^2$, it appears with only one derivative in $\ln k$. Hence, tensor corrections are as important as scalar corrections in slow-roll inflation.

The remaining task is to average the background quantities, given in eq.~\eqref{eq:geometry_definition_x_dep_background}. In principle, we have to average $\zbar(\vec{x})$ and $\gbar(\vec{x})$ over the large observed region of box-size $L$. However, this is equivalent to an ensemble average of $\zbar(0)$ and $\gbar(0)$ with IR cut-off $L$. Thus, in single-field, slow-roll inflation, we are dealing with sums of Gaussian random variables $\random{q}$, respectively $\randtens{q}{s}$,
\begin{align}
  \label{eq:geometry_definition_zbar}
  \zbar  &= \int\limits_{1/L}^k \kmes{q} \; \zeta(\qv) = \int\limits_{1/L}^k \kmes{q} \; \frac{\None (\qv) \, H(\qv)}{\sqrt{2 q^3}} \; \random{q} \\
  \label{eq:geometry_definition_gbar}
  \gbar_{ij}  &= \int\limits_{1/L}^k \kmes{q} \; \gamma_{ij}(\qv) = \int\limits_{1/L}^k \kmes{q} \;\sum\limits_{s=+,\times} \frac{H(\qv)}{\sqrt{q^3}} \; \epsilon^s_{ij}(\kv) \; \randtens{q}{s} \quad .
\end{align}
While their averages are vanishing, $\corrbase{\zbar} = 0 = \corrbase{\gbar_{ij}}$, one finds a scale-dependent result for the two-point functions. For instance under the assumption of a scale-invariant power spectrum, they obey a logarithmic scale-dependence 
\begin{align}
\label{eq:correlators_zbar_gbar_squared}
 \corrbase{\zbar^2} &= \left( \frac{\None H}{2\pi} \right)^2 \ln(kL) & \corrbase{\tr \gbar^2} = \corr{\gbar_{ij}}{\gbar_{ij}} = 8 \left(\frac{H}{2\pi} \right)^2 \ln(kL) \;\;\; .
\end{align}
Neglecting tensor fluctuations for the moment, the background $\zbar$ is  a sum of Gaussian random variables $\random{q}$. Therefore, $\zbar$ itself is a Gaussian random variable, with distribution\footnote{This is related to the stochastic approach \cite{Starobinsky:1986fx} of Starobinsky.}
\be\label{gaussiandistr}
{\mathbb P}\left[\bar \zeta\right] d \bar\zeta
\,=\,  \frac{1}{\sqrt{2\pi \sigma_\zeta^2}} \;  \exp\left(-\frac{\zbar^2}{2\sigma_\zeta^2}\right) d \bar\zeta
\ee
where the width is
\begin{align}
  \sigma^2_\zeta &= \corrbase{\zbar^2} = \int\limits_{1/L}^k \kmes{q} \; \frac{\None^2 (\qv) \, H^2(\qv)}{2 q^3}  \qquad .
\end{align}
Note that we do not assume a scale-invariant behavior of the power spectrum in this expression. Typically, the $n$-point functions we are interested in can be expressed as $\corrbase{f(\,\zbar(\xv)\,)}$, for some function $f$. As usual for Gaussian variables, this may be expressed in terms of an integral over a Gaussian probability distribution
\be
  \corrbase{f(\zbar)} = \frac{1}{\sqrt{2\pi \sigma_\zeta^2}} \; \int d\zbar \; \exp\left(-\frac{\zbar^2}{2\sigma_\zeta^2}\right) \; f(\zbar)  \qquad .
\ee
As an example, the power spectrum is given by
\be
  \label{eq:geometry_power_integral}
  \powze (k) = \frac{1}{\sqrt{2\pi \sigma_\zeta^2}} \; \int d\zbar \; \exp\left(-\frac{\zbar^2}{2\sigma_\zeta^2}\right) \; \treepowze (k e^{-\zbar})  \qquad .
\ee
Consequently, the question of convergence of fluctuations due to long-wavelength modes reduces to convergence properties of this single integral. The usual series expansion can be recovered by expanding the function $\treepowze$ in the logarithm of the scale $k$. This yields
\begin{align}
 \label{eq:geometry_recover_series}
 \powze (k) &= \sum\limits_{n=0}^\infty \; \frac{\corrbase{\zbar^{2n}}}{(2n)!} \; \frac{d^{2n} \, \treepowze (k)}{d(\ln k)^{2n}}  \\
 \corrbase{\zbar^{2n}} & = \frac{1}{\sqrt{2\pi \sigma_\zeta^2}} \int d\zbar \; \zbar^{2n} \; \exp\left(-\frac{\zbar^2}{2\sigma_\zeta^2}\right)\,=\,\left(2 n -1\right)!! \,\left(  \sigma_\zeta^2\right)^n \quad ,
 \label{eq:geometry_zbarton_in_zbar_squared}
\end{align}
where $n!!$ denotes the double factorial. This is in agreement with \cite{Giddings:2010nc}. We emphasize, however, that a breakdown of convergence of the series does not necessarily mean a breakdown of convergence of the integral in eq.~\eqref{eq:geometry_power_integral}. We return to this point in sect.~\ref{sec:power_spectrum}. Notice also that  
in eqs.~\eqref{eq:geometry_definition_zbar} and \eqref{eq:geometry_definition_gbar} we have neglected the intrinsic non-Gaussianity of curvature and tensor perturbations. 
Such intrinsic non-Gaussianity is  present at sub-leading order in slow-roll. However, at every log-order, there is a term consisting solely of Gaussian contributions. Relative to this term, contributions with intrinsic non-Gaussian parts are suppressed by slow-roll parameters and the Hubble scale with no additional log-enhancement. Therefore, neglecting the non-Gaussian contribution is justified in our leading-log analysis of IR-corrections.

Attention has to be paid to the fact that inflation has ended at some point. Hence, there exists a value $k_{\rm max}$ corresponding to modes that have never left the horizon. 
The observer measuring $\powze(k)$ for some fixed $k$ will have to exclude regions where $k e^{-\zbar} > k_{\rm max}$ from his averaging procedure. Technically, this implies a lower bound for the $\zbar$-integral, given by $\zbar_{\rm min} = -\ln(k_{\rm max} / k)$.
\be
  \label{eq:geometry_all_order_scalar_integral}
  \powze (k) = \frac{1}{\sqrt{2\pi \sigma_\zeta^2}} \; \int\limits_{ \zbar_{\rm min} }^\infty d\zbar \; \exp\left(-\frac{\zbar^2}{2\sigma_\zeta^2}\right) \; \treepowze (k e^{-\zbar})  \qquad .
\ee
The Gaussian function in eq.~\eqref{eq:geometry_all_order_scalar_integral} gives a non-negligible contribution only in a limited range around zero. This range is of the order of $\sigma_\zeta$. For large $L$ (implying large $\sigma_\zeta$) and for $k$ sufficiently close to $k_{\rm max}$, the lower bound $\zbar_{\rm min}$ enters this range. Hence, in such cases, the lower bound implies the subtraction of a significant contribution from the integral. We finally note that the existence of $k_{\rm max}$ and $\zbar_{\rm min}$ are related to potential convergence problems of the series expansion in eq.~\eqref{eq:geometry_recover_series}. This is apparent since the slow-roll conditions, which are responsible for the smallness of derivatives of $\powze (k)$, break down near $k_{\rm max}$.

\smallskip

Including tensor modes is in principle straightforward, but complicated by the matrix structure of $\gbar$ and the different independent polarizations involved. In order not to overburden formulae, we set the scalar background $\zbar$ to zero in what follows. The complete power spectrum can then be expressed as
\be
  \powze (k) = \left\langle \; \left[  \left(e^{-\gbar }\right)_{ij} \hat{k}_i\hat{k}_j \right]^{-3/2}  \; \treepowze 
\left(\left[
  \left(e^{-\gbar }\right)_{ij} \hat{k}_i\hat{k}_j \right]^{\frac12} k \,
\right)  \; \right\rangle \qquad ,
\ee
where $\hat k$ is the unit vector parallel to $\vec k$. We also introduce the notation
\be
 n\,\equiv\,\left[  \left(e^{-\gbar }\right)_{lm} \hat{k}_l\hat{k}_m \right] \;\; .
\ee
Note that each entry of the matrix $\bar{\gamma}_{ij}$, being a sum of Gaussian random variables, is a Gaussian random variable. However, the various entries in the matrix are not statistically independent: this implies that it is not obvious how to calculate the statistical distribution  of the entries of the {\it exponential} of $(-\bar \gamma)$, that enters in the definition of $n$. Having this distribution, that we denote with ${\mathbb P}\left[  n \right]$, it is straightforward to provide an integral representation for the power spectrum subject to tensor background modes:
\be
\powze (k) = \int d n \,\,\mathbb{P} \left[ n \right] \,\, n^{-\frac32}(\hat k) \treepowze  \left( n^{\frac12}\,k \right)  \;\; .
\ee
It is clear that, at least numerically, $\mathbb{P} \left[ n \right]$ can be determined and the integral can be calculated.

\subsection{Higher correlation functions}

To discuss $n$-point functions, we could try to generalize the 'almost scale-invariant' spectrum of eq.~\eqref{eq:geometry_power_spectrum} by writing
\be
 \mathcal{P}_{n} (\vec{k}_1,\ldots ,\vec{k}_n)  = \left\langle \; \left( \frac{k^3}{2\pi^2} \right)^n \int\limits d^3 y_1 \ldots d^3 y_n \; e^{-i(\vec{k}_1\vec{y}_1 + \ldots + \vec{k}_n\vec{y}_n)} \, \zeta(\xv) \; \zeta(\xv + \vec{y}_1) \ldots \zeta(\xv + \vec{y}_n)  \; \right\rangle   \quad .
\ee
However, it is not clear which particular combination of $ k_1 \ldots k_n$ one should use to define $k$ in the prefactor $k^{3n}$. This is not irrelevant since factors $e^{\gbar/2}$ will get tangled up in this prefactor. Hence, we choose to write the general formula for the higher-order analogue of the conventional spectrum $P_\zeta (k) = 2 \pi^2 \powze (k) / k^3$. In doing so, prefactors will arise from the scaling of the $d^3y_a$. Since the determinant of the tensor contribution is one, this scaling consist exclusively of $\zbar$, which only depend on the overall scale. Given these preliminaries, the generalization of our formalism is completely straightforward and the IR-safe spectrum is defined as
\be
  \treenpow{n} (\vec{k}_1, \ldots , \vec{k}_n) = \left\langle \; \int\limits d^3 z_1 \ldots d^3 z_n \; e^{-i(\vec{k}_1\vec{z}_1 + \ldots + \vec{k}_n\vec{z}_n)} \, \zeta(\xv) \; \zeta(\xv + \vec{y}_1) \ldots \zeta(\xv + \vec{y}_n)  \; \right\rangle \quad ,
\ee
where
\be
  \vec{y}_a = \vec{y}_a (\vec{z}_a, \zbar, \gbar) = e^{-\zbar}  e^{-\gbar/2} \vec{z}_a \qquad .
\ee
This means that we measure the correlation function in terms of $n$ invariant distances, characterized by a set of vectors $\vec{z}_a$, $a\in [1,\ldots, n]$. Hence, the $\vec{z}_a$-dependence of the corresponding $n$-point function is independent of background quantities and, therefore, IR-safe. Consequently, its Fourier transform, i.e.\ the spectrum $\treenpow{n}$, is the desired IR-safe spectrum. A straightforward generalization of the previous calculation for the power spectrum provides the following result
\be
  \label{eq:geometry_comparison_higher_spectra}
  \npow{n} (\vec{k}_1, \ldots , \vec{k}_n) = \left\langle  e^{-3n\zbar} \, \treenpow{n} ( e^{-\zbar} e^{-\gbar/2} \vec{k}_1 , \ldots , e^{-\zbar} e^{-\gbar/2} \vec{k}_n) \right\rangle   \qquad .
\ee
As already stressed above, the prefactor $e^{-3n\zbar}$ originates from the naive scaling $\treenpow{n} \sim k^{-3n}$. 

The log-enhancement-effects of higher correlation functions specified by eq.~\eqref{eq:geometry_comparison_higher_spectra} can be directly applied to observables measuring non-Gaussianity, like $f_{\rm NL}$, as we are going to discuss in section \ref{sec:3_point_bispectrum}.

\section{An alternative approach within slow-roll inflation}

\label{sec:alternative_way}

In the previous section we discussed a systematic way to define IR-safe $n$-point functions. We have explained how to straightforwardly obtain, from these IR-safe quantities, the  corresponding IR-sensitive objects. In this section, we present an alternative point of view: working only in momentum space, we will directly calculate the all-orders IR-enhancement of the conventional power spectrum. To be more specific, we will compute the curvature perturbation $\zeta$, by implementing a suitable extension of the $\delta N$-formalism, in such a way as to include the effects of long-wavelength modes. The results coincide with what we obtained in the previous section, in all cases in which $\delta N$-formalism is applicable. So, in these cases, the two methods are equivalent.
\smallskip

We focus on a single, slowly rolling  scalar field $\phi$ (the extension to multiple fields is outlined in Appendix \ref{AppA}). We assume the underlying metric to be of the form
\be
  ds^2 = -dt^2 + a^2(t)\,\bar{g}_{ij} dx^i dx^j \qquad .
\ee
Throughout this section, we are interested in quantities evaluated at some wave vector $\kv$. To analyze contributions from the background to these quantities, we find it technically convenient to separate the  fluctuations into modes characterized by momenta larger, and smaller, than $k$. For modes $\qv$ with $q=|\qv| \ll k$, we will work in a gauge with $\delta_{\qv} \phi=0$. While for $q$ around $k$ and larger, we adopt a gauge with vanishing scalar metric fluctuations. The advantage of this splitting, and of these different gauge choices, is that the  contribution from long-wavelength modes is contained in geometrical quantities and, therefore, contained in the 3-metric
\be
  \label{eq:alternative_definition_3_metric}
  \bar{g}_{ij} = e^{2\zbar} \left( e^{\gbar} \right)_{ij} \quad .
\ee
Here, the scalar and tensor background, $\zbar$ and $\gbar_{ij}$, are defined as before. It would be interesting to understand whether the above construction
can be done in a gauge invariant manner. On the other hand, let us stress that
we proceed in this way only for technical convenience. One could also work
with a gauge characterized by vanishing scalar metric fluctuations for all $\qv$. With this choice, however, the scalar background from long-wavelength $\delta_{\qv} \phi$ would affect the scalar field value at the time of horizon exit (see \cite{Byrnes:2010yc} for a treatment of background modes of the scalar field $\phi$ in this latter gauge choice). In contrast, the tensor background would still enter via the 3-metric. Therefore, the inclusion of tensor background modes within the $\delta N$-formalism requires a treatment as outlined in this section, contrary to the scalar background which might be calculated by different techniques.

The appearance of background contributions in the 3-metric eq.~\eqref{eq:alternative_definition_3_metric} has important consequences for the physical length scale associated with the wave vector $\kv$, i.e.\ on the physical wavelength. Due to the deviation of $\bar{g}_{ij}$ from flatness, this scale is not the inverse of $k=\sqrt{k_i k_j \delta_{ij}}$, but is instead given by $1/k'$ with
\be
  \label{eq:alternative_definition_k_prime} 
  k'^2 \,= \,e^{-2\zbar} \,\left( e^{- \gbar}\right)_{ij} \, k_i  k_j \qquad .
\ee
Hence, the physical scale depends on the original vector $\kv$ and on the background quantities $\bar \zeta$ and $\bar \gamma_{ij}$. This dependence on background quantities leads to a shift in the time of horizon exit for a given scalar mode of momentum $\kv$, from $t_k$ to $t_{k'}$. That is, since the time of horizon exit is defined by the relation $k\,=\,a(t_k) H(t_k)$ ($a$ being the scale factor), at first order in slow-roll we have the relation
\be
  dt_k \, = \, \frac{1}{H} \, d\ln{k} \qquad .
\ee
For small time variations, and at leading order in slow-roll, we can integrate the previous equation and find
\begin{align}
  \label{eq:alternative_shift_Hubble}
  H \left( t_{k'}-t_{k}\right) &= \,\ln{\frac{k'}{k}} \, = \,-\zbar - \Delta \qquad {\rm with} \\
  \Delta  & \equiv \left( \frac{1}{2} \gbar_{ij} - \frac{1}{4} \gbar_{il} \gbar_{lj} \right) \hat{k}_i \hat{k}_j + \left( \frac{1}{2} \gbar_{ij} \; \hat{k}_i \hat{k}_j \right)^2 + \mathcal{O}(\gbar^3)
  \label{eq:alternative_def_tensors_shift_Delta} \qquad .
\end{align}
Here, $\hat{k}$ represents a unit vector in $\kv$-direction. The quantity $\Delta$ collects the leading order contributions from the long-wavelength tensor modes, obtained from expanding the exponential in eq.~\eqref{eq:alternative_definition_k_prime}.
\smallskip

The form of the background metric affects the dynamics of first order, massless scalar fluctuations. In momentum space, the equation of motion for the scalar perturbations reads 
\be
  \label{eq:alternative_eom}
  \left( \delta_{\kv} \phi\right)^{..} + 3 H \left( \delta_{\kv} \phi\right)^{.} + 
\frac{k'^2}{a^2} \, \delta_{\kv} \phi \, = \, 0  \qquad ,
\ee
where dots denote derivatives with respect to time. Note that the effect of background quantities enters via the Laplacian which leads to the $k'^2$ instead of $k^2$ in the third term on the left-hand side. The solution for the fluctuation $\delta_{\vec k} \phi$ results in
\be
 \label{eq:alternative_solution_dphi}
 \delta_{\kv}\phi  = \delta_{\kv}\phi (k',  \bar g_{ij}) = \frac{H(k')}{\left({\rm det}^{\frac14}\, \bar g_{ij}\right) \, \left(2 k'^3\right)^{\frac12}} \; \random{k}
\ee
in a superhorizon regime. In our notation $H(k')$ indicates that this quantity is evaluated at time of horizon exit of the scale $k'$, instead of $k$, in order to take into account the shift due to long-wavelength contributions. The normalization of $\delta_{\kv} \phi$, ${\rm det}^{1/4}\, \bar g_{ij}$ in the denominator, is obtained when imposing the usual commutation relations between the quantized scalar fluctuation and its momentum conjugate (see, for example, \cite{Ford:1997hb}). Another way to understand it is the following: the normalization of $\delta_{\kv} \phi$ is set by requiring that in the limit of short distances $y$, we recover the singularity of the scalar field in Minkowski space for $\corr{\delta\phi(\xv)}{\delta\phi(\xv+\yv)}$. On these distances, the background quantities $\zbar$ and $\gbar_{ij}$ are constant and, hence, need to be absorbed in a redefinition of space variables in order to bring the metric in Minkowski-form. This redefinition is responsible for the factor ${\rm det}^{1/4}\, \bar g_{ij}$, appearing in the normalization of $\delta_{\kv} \phi$.

Since ${\bar \gamma}_{ij}$ is traceless, ${\rm det}\, \bar g_{ij}=\exp(6 \bar{\zeta})$  and eq.~\eqref{eq:alternative_solution_dphi} can be rewritten as
\begin{align}
  \delta_{\kv} \phi(k',  \bar g_{ij}) &= \,\frac{H(k')}{\sqrt{2}\left[ k_i k_j \left(e^{-\bar \gamma}\right)_{ij}\right]^{\frac34}} \; \random{k} \nonumber \\ 
  &= m^{\frac12}(\hat{k})\,\frac{H(k')}{\left(2 k^3  \right)^{\frac12}} \; \random{k}  \quad ,
  \label{newflu}
\end{align}
where we define the function $m(\hat{k})$ that depends on a unit
vector $\hat k$ along the direction of $\kv$: 
\be
  \label{eq:alternative_definition_m}
  m(\hat{k})\,\equiv\,\left[\left( e^{-\gbar}\right)_{ij} \hat{k}_i \hat{k}_j \right]^{-\frac32} \qquad .
\ee
In equation~\eqref{newflu}, the dependence on background quantities is limited to the overall function $m(\hat{k})$ (that depends only on the tensor background, see eq.~\eqref{eq:alternative_definition_m}) and to the `time' argument $k'$ of the Hubble parameter.
\smallskip

Starting from scalar fluctuations and by using $\delta N$-formalism \cite{Lyth:2005fi, Starobinsky:1986fxa, Sasaki:1995aw, Wands:2000dp, Lyth:2004gb}, we can express the curvature fluctuation $\zeta$ at superhorizon scales on a constant energy density slice, that we take to be the reheating surface, in terms of $\delta\phi$. The curvature perturbation $\zeta_{\vec k}$ is related to the time integral of the local expansion parameter, providing the number of e-foldings, from an initial hypersurface (that we take at time of horizon exit for the mode $\vec k$) to  the final hypersurface of constant energy density. In single field inflation, we have
\be
 \zeta\,=\,N\left[\phi+\delta \phi\right]-\langle N \rangle  \qquad ,
\ee
where $\langle N \rangle$ is the spatial average of the first term on the right-hand side. The quantity $\phi+\delta \phi$ corresponds to the homogeneous value for the scalar field plus its perturbation built, as above, on a space-time geometry that includes the contributions of long-wavelength modes. The previous schematic expression can be expanded in the scalar fluctuations, and gives in momentum space~\footnote{For the purposes of this work, we can truncate the $\delta N$ expansion to the first, leading order term in slow-roll. Including higher order terms is straightforward, as we discuss in Appendix \ref{AppA}.}
\be
  \label{fdze}
  \zeta_{\kv}\,=\,\None (k') \,\delta_{\kv} \phi(k', \bar g_{ij})+\dots \qquad .
\ee
Notice that functions on the right-hand-side are evaluated at time of horizon exit of the mode $\vec{k}$, which is sensitive to the change in the background geometry due to long-wavelength modes. That is, their  argument is $k'$ instead of $k$. As in section \ref{sec:geometry_of_the_reheating_surface}, the function $N_\phi = dN/d\phi$ is given by $\None = V/(dV/d\phi)$. The remaining terms in the $\delta N$ expansion, understood in the dots of eq.~\eqref{fdze}, are slow-roll suppressed with respect to the first one. Using the results obtained earlier, we get for $\zeta_{\vec k}$ an expression in terms of Gaussian random variables as follows
\be
 \label{eq:alternative_def_zeta}
 \zeta_{\kv} = \left[ m^{\frac12}(\hat{k}) \, \None(k')\,H(k') \right] \; \frac{\random{k}}{\left(2 k^3  \right)^{\frac12}} \quad .
\ee
The dependence on long-wavelength background quantities is contained in the overall factor between squared parenthesis. Eq.~\eqref{eq:alternative_def_zeta}, possibly including higher order terms in the $\delta N$-expansion, is all what we need to straightforwardly compute inflationary observables, associated to $n$-point functions of curvature perturbations, including the effects of long-wavelength modes. Eq.~\eqref{eq:alternative_def_zeta} can be regarded as an extension of $\delta N$-formalism. It includes the contributions of long-wavelength scalar and tensor modes in the expression for the curvature perturbation $\zeta$.
\smallskip

As an application of eq.~\eqref{eq:alternative_def_zeta}, we rederive the expression for log-enhanced contributions to the power spectrum. We start with the two-point function of the curvature perturbation
\bea
  \corr{\zeta_{\vec{k}}}{\zeta_{\vec{p}}} &=&\frac{1}{2\left(k p  \right)^{\frac32} } \; \langle \; m^\frac12(\hat k) m^\frac12(\hat p)\, N_{\phi}(k')H(k')\,N_{\phi}(p') H(p')
  \; \random{k} \random{p} \; \rangle \nonumber \\ 
&=&\frac{(2\pi)^3\delta^{(3)}(\vec{k}+\vec{p})
}{2 k^{3} 
}\; \langle \,
 m(\hat k) \,N_\phi^2(k') H^2(k') \, \rangle \qquad ,
\eea
where for passing from first to second line, we used Wick's theorem and contracted the Gaussian variables $\random{k}$ and $\random{p}$. Indeed, $\random{k}$ is only allowed to contract with $\random{p}$, since any other quantity depends on modes with momenta much smaller than $k$. One obtains the following expression for the power spectrum:
\be\label{copos}
  \powze (k)\,=\,\frac{1}{(2\pi)^2}\langle
   m(\hat k) \,N_\phi^2(k') H^2(k')
\rangle  \qquad .
\ee
The argument of the average on the right-hand side depends on long-wavelength scalar and tensor contributions, which, as shown in eq.~\eqref{eq:correlators_zbar_gbar_squared}, have non-vanishing two-point functions. In absence of contributions of long-wavelength modes, eq. (\ref{copos}) provides the following tree-level result
\be
 \treepowze (k)\,=\,\frac{1}{(2\pi)^2} \, N_\phi^2(k) H^2(k) \qquad ,
\ee
which also coincides with the definition of the IR-safe power spectrum provided in section \ref{sec:geometry_of_the_reheating_surface}. Notice that the dependence on the scale $k$ in $\treepowze$ occurs only through the dependence on the time of horizon exit of the right-hand side. Using this fact, eq.~\eqref{copos} can be rewritten as
\be
 \label{copos2}
 \powze (k)\,=\,\langle m(\hat{k}) \treepowze (k')\rangle \quad .
\ee
Recall that $\hat k$ represents the unit vector along the direction of $\vec k$, while $k'$ in the previous expression is associated to $k$ via eq.~\eqref{eq:alternative_definition_k_prime}. Using these formulae, eq.~\eqref{copos2} can be rewritten as
\be
  \powze (k) = \left\langle \; \left[  \left(e^{-\gbar}\right)_{ij} \hat{k}_i\hat{k}_j \right]^{-3/2}  \; \treepowze \left( e^{-\zbar} e^{-\gbar/2}  \kv \,\right)  \; \right\rangle \qquad .
\ee
Not surprisingly, this corresponds exactly to equation \eqref{eq:geometry_comparison_power_spectra}, obtained with the method of section \ref{sec:geometry_of_the_reheating_surface}.

\section{Two-point function and the power spectrum}

\label{sec:power_spectrum}

In this section, we analyze the log-enhanced corrections due to scalar and tensor long-wavelength modes to the power spectrum of curvature perturbation. Using the results from sec.~\ref{sec:geometry_of_the_reheating_surface} or sec.~\ref{sec:alternative_way}, the power spectrum is given by the formula
\begin{align}
  \powze (k) &= \langle m(\hat{k}) {\cal P}^{(0)}_{\zeta}(k')\rangle\label{tptea}\\
             &= \left\langle \; \left[  \left(e^{-\gbar}\right)_{ij} \hat{k}_i\hat{k}_j \right]^{-3/2}  \; \treepowze \left( e^{-\zbar} e^{-\gbar/2}  \kv \,\right)  \; \right\rangle
             \label{eq:two_point_comparison_power_spectra}\qquad .
\end{align}
As explained in section \ref{sec:geometry_of_the_reheating_surface}, this implies that we can deal with scalar perturbations to all orders, resumming the complete series, in case an exact expression for the tree level power spectrum is known. We will return to this important topic at the end of this section; for the moment we focus on calculating, in full generality, the leading log-enhanced contributions to the power spectrum. In order to do so, it is sufficient to expand eq.~\eqref{eq:two_point_comparison_power_spectra} in $\zbar$ and $\gbar$. The following equations are useful for this purpose
\begin{align}
 m(\hat{k}) &=  \left[  \left(e^{-\gbar}\right)_{ij} \hat{k}_i\hat{k}_j \right]^{-3/2} = 1 + \frac{3}{2} \gbar_{ij} \, \hat{k}_i \hat{k}_j - \frac{3}{4} \gbar_{il} \gbar_{lj} \, \hat{k}_i \hat{k}_j + \frac{15}{8} \left( \gbar_{ij} \, \hat{k}_i \hat{k}_j \right)^2 + \mathcal{O}(\gbar_{ij}^3)  \\
 \ln k' &= \ln k -\zbar - \Delta \\
 \Delta &= \frac{1}{2} \, \gbar_{ij} \hat{k}_i\hat{k}_j - \frac{1}{4} \gbar_{il} \gbar_{lj} \, \hat{k}_i\hat{k}_j + \frac{1}{4} \left(\gbar_{ij} \, \hat{k}_i\hat{k}_j \right)^2 + \mathcal{O}(\gbar_{ij}^3) \qquad .
\end{align}
Here, $k'$ denotes the Euclidean length of the vector $e^{-\zbar} e^{-\gbar/2}  \kv$. We will also make use of the identity (see also \cite{Giddings:2010nc}):
\be
  \corr{\gbar_{ij}}{\gbar_{lm}} = \frac{1}{30} \,\corrbase{\tr\gbar^2}\, \left[ \, 3 \left(  \delta_{il}\delta_{jm} +  \delta_{im}\delta_{jl} \right) - 2 \,  \delta_{ij}\delta_{lm}\right]  \qquad ,
\ee
where $\corrbase{\tr\gbar^2} = \sum_{ij} \corr{\gbar_{ij}}{\gbar_{ij}}$. From this, it is easy to check that a cancellation leads to $\langle m({\hat{k}})\rangle=1+ \mathcal{O}(\gbar^4)$. We can then expand eq.~\eqref{tptea} up to the first non-vanishing contributions. We obtain
\begin{align}
 \powze (k) &=\,\treepowze (k) \, \left\langle m(\hat{k})  \cdot\left[1-\left(\Delta + \zbar \right) \; \frac{1}{\treepowze(k)} \, \frac{d\treepowze(k)}{d\ln k} + \frac{\zbar^2}{2} \; \frac{1}{\treepowze(k)} \frac{d^2\treepowze(k)}{d(\ln k)^2} \right] \right\rangle   \nonumber\\
  & = \, \left\{ 1 -\left[ \corrbase{ (m(\hat{k})-1) \, \Delta } +  \corrbase{\Delta} \right]  \, \, \frac{d}{d\ln k} + \frac{\corrbase{\zbar^2}}{2}  \frac{d^2}{d(\ln k)^2} \right\}\treepowze (k)  \nonumber \\
   & =  \left( 1 - \frac{1}{20} \corrbase{{\rm tr} \,\gbar^2} \frac{d}{d\ln k}  + \frac{1}{2} \corrbase{\zbar^2} \frac{d^2}{d(\ln k)^2} \right) \treepowze (k) \qquad .
 \label{fiveps}
\end{align}
This equation was also found in \cite{Giddings:2010nc}~\footnote{Note that in \cite{Giddings:2010nc} the result for corrections due to tensors is expressed in terms of the quantity $\corrbase{\gbar^2_{\rm GS}} \equiv \frac{1}{4} \sum_{ij} \corr{\gbar_{ij}}{\gbar_{ij}} = \frac{1}{4} \corrbase{\tr\gbar^2}$~. Tensor contributions to inflationary observables, using different methods, have been also considered in \cite{Dimastrogiovanni:2008af,Seery:2008ax}.}. Neglecting tensor contributions, this reproduces the results of log-enhanced corrections to the power spectrum due to scalar fluctuations given in \cite{Byrnes:2010yc, senatore-talk}.

Taking another point of view, we note that eq.~\eqref{fiveps} can be obtained 
by expanding $\zeta$, the curvature perturbation in uniform-energy-density 
gauge, in terms of $\delta\phi$, the scalar field perturbation in flat gauge. 
Up to quadratic order in $\delta\phi$, the relevant gauge transformation 
is the one between $\zeta$ and $\zeta_n=-(H/\dot{\phi})\delta\phi$ given 
in eq.~(A.8) of \cite{Maldacena:2002vr}. If we focus on terms that are 
leading order in slow-roll and neglect terms vanishing at superhorizon 
scales, all IR divergences arising from the expansion in (A.8) of 
\cite{Maldacena:2002vr} are captured by our result. However, the complete 
IR correction requires the inclusion of term $\sim\delta\phi^3$. This can be 
realized using the $\delta N$ formalism, and it was shown in 
\cite{Byrnes:2010yc} that an appropriately modified version 
of this formalism correctly computes the scalar part of eq.~\eqref{fiveps}
(see also Sect.~\ref{sec:alternative_way} of the present paper). 

For a weakly scale-dependent power spectrum, the explicit values for $\corrbase{\zbar^2}$ and $\corrbase{\tr\gbar^2}$ were already given in eq.~\eqref{eq:correlators_zbar_gbar_squared}. Using the definitions of the spectral index of curvature perturbations $n_\zeta$, its running $\alpha_\zeta$ and the tensor-to-scalar ratio $r=\None^{-2}= 2\epsilon$, 
\begin{align}
 \label{eq:two_point_def_specindex_running_tensortoscalar}
 n_\zeta -1 &= \frac{d\ln\treepowze}{d\ln k}   &
 \alpha_\zeta & = \frac{d^2\ln\treepowze}{d(\ln k)^2}   &
 r &= \frac{\corrbase{\tr\gbar^2}}{8\corrbase{\zbar^2}} \quad ,
\end{align}
the leading order correction to the power spectrum can be written as
\be
  \label{eq:two_point_power_spectrum_in_observables}
  \powze (k) \,=\, \, \treepowze (k) \, \left\{ 1 +\frac12 \left[ (n_\zeta -1)^2 + \alpha_\zeta -\frac{4 r}{5} \, (n_\zeta -1) \right]\treepowze (k)\,\ln(kL) \right\}
\ee
The agreement\footnote{Note that the authors of \cite{Giddings:2010nc} chose a different parameterization of the power spectrum, namely $\treepowze \sim k^{n(k)-1}$. This leads to slightly different numerical factors. For instance, $d^2\treepowze/d(\ln k)^2 = [(n-1)^2 + 2\alpha]\treepowze$  in their parameterization.} of eqs.~\eqref{fiveps} and \eqref{eq:two_point_power_spectrum_in_observables} with \cite{Giddings:2010nc} is a non-trivial check for our approach.

We learn that long-wavelength modes provide log-enhanced contributions to the power-spectrum that are suppressed by second order slow-roll parameters, and by a factor of $\treepowze$. The latter is determined by WMAP to be $\treepowze \simeq 2.3\times 10^{-9}$ \cite{Komatsu:2010fb}.

Let us stress the quite non-trivial fact that scalar and tensor long-wavelength modes contribute at the same (second) order in a slow-roll expansion. This is  due to the cancellation leading to  $\langle m(\hat k) \rangle=1+\mathcal{O}(\gbar^4)$. In any case, this property is specific of the power spectrum: as we will learn in the next section, corrections to non-Gaussianity parameters do not share this property.

\bigskip

Having calculated the leading order correction to the power spectrum, we turn to evaluating scalar perturbations to all orders as described in section \ref{sec:geometry_of_the_reheating_surface}. We start with a inflationary potential for which the spectral index is constant. This is realized for the famous example of power law inflation \cite{Lucchin:1984yf}. The potential is $V=V_0 \exp{\left[ - \sqrt{\frac{2}{q}}\,\phi\right]}$,  with constant $q$, and the scale factor evolves as $a(t) =  a_0 t^q$. In this set-up, the equations for scalar fluctuations can be solved exactly without having to rely on a slow-roll approximation. For this particular model, we assume that our scale of interest $k$ is much smaller than $k_{\rm max}$, reflecting the transition to scales that have never left the horizon during inflation. Hence, the integral in eq.~\eqref{eq:geometry_all_order_scalar_integral} is well approximated by setting $\zbar_{\rm min}$ to $-\infty$. The power spectrum of curvature perturbations reads \cite{Lyth:1991bc,Stewart:1993bc}
\bea
 {\cal{P}}^{(0)}_{\zeta} (k) &=& \treepowze (k_0) 
\left( \frac{k}{k_0} \right)^{-2/(q-1)
} \quad .
\eea
So the spectral index $n_\zeta -1=-2/(q-1)$ is constant as desired. We then obtain
\be
  \label{funforin}
  \treepowze (k e^{-\bar \zeta}) = \treepowze (k)\,\, e^{-(n_\zeta -1)\bar \zeta} \quad .
\ee
Plugging this expression into eq.~\eqref{eq:geometry_all_order_scalar_integral}, one finds an integral that can be solved analytically. We get
\be
  \powze (k) =  \treepowze (k)
\exp\left( \frac{ \sigma^2_\zeta (n_\zeta -1)^2}{2 
} \right)  \qquad .
\ee
This expression captures {\it at all orders} the contributions of long-wavelength modes. In a sense, we are providing the function whose series expansion has been found in \cite{Giddings:2010nc}. Notice that the corrections are not independent of the scale $k$ since $\sigma^2_\zeta$ is a function of $k$ (e.g.\  $\sigma^2_\zeta= \treepowze \; \ln(kL)$ for a weak scale-dependence of $\treepowze$).

We expect a similar behavior including other contributions in more general models of inflation, for example associated with the running of the spectral index; however, solving the integral analytically might be more difficult in these models. For instance, the chaotic potential investigated in \cite{Giddings:2010nc}, $V(\phi) = \lambda \phi^\alpha$ $(\alpha >0)$, leads to the following tree-level power spectrum:
\be
  \label{eq:2_point_monomial_potential_tree_power_spectrum}
  \treepowze(k) = \left( \frac{\None H}{2\pi} \right)^2 = \frac{1}{(2\pi)^2} \frac{\lambda}{3\alpha^2} \phi^{\alpha+2} (k) \quad .
\ee
The scalar field value in dependence of the horizon-exit time of the mode $k$ is given by the differential equation $d\phi/(d\ln k) = -\alpha/\phi$. This can be integrated to yield
\be
  \phi(k) = \sqrt{\phi^2(k_{\rm max}) + 2\alpha \ln \frac{k_{\rm max}}{k} } \qquad .
\ee
Note that the condition $\zbar \geq \zbar_{\rm min}$ guarantees $\phi( e^{-\zbar} k ) \geq \phi(k_{\rm max})$ for all possible values $\zbar$. Hence, the integral in eq.~\eqref{eq:geometry_all_order_scalar_integral} is well-defined and finite. As already described in sec.~\ref{sec:geometry_of_the_reheating_surface}, the series expansion can be recovered easily from the integral expression in eq.~\eqref{eq:geometry_all_order_scalar_integral}.  For this purpose, one can expand $\treepowze ( e^{-\zbar} k )$, as given in eq.~\eqref{eq:2_point_monomial_potential_tree_power_spectrum}, and make use of the moments for the Gaussian probability distribution (see eqs.~\eqref{eq:geometry_recover_series} and \eqref{eq:geometry_zbarton_in_zbar_squared}~). Derivatives of the power spectrum \eqref{eq:2_point_monomial_potential_tree_power_spectrum} w.r.t.\ $\ln k$ can be expressed in terms of the spectral index $n_\zeta -1 = -\alpha (\alpha +2) / \phi^2$ and the model parameter $\alpha$ :
\be
  A_l = \frac{1}{\treepowze} \, \frac{d^l \treepowze}{d(\ln k)^l} = \left( \frac{n_\zeta -1}{\frac{\alpha}{2}+1} \right)^l \prod\limits_{i=1}^l (\frac{\alpha}{2}+2-i)  \;\; .
\ee
Hence, the series expansion is given by
\be
  \powze (k) = \treepowze (k) \left[  1 + \sum\limits_{n=1}^\infty \frac{(2n-1)!!}{(2n)!} \; A_{2n} \;\left(\sigma_\zeta^2 \right)^n  \right]  \;\; .
\ee

\bigskip

In the last part of this section, we discuss the question of convergence of the series expansion returning to the general case (see also \cite{Giddings:2010nc}). The series expansion in eq.~\eqref{eq:geometry_recover_series} was
\begin{align}
 \label{eq:2_point_series_expansion}
 \powze (k) &= \treepowze (k) \left[ 1+ \sum\limits_{n=1}^\infty \; \frac{\corrbase{\zbar^{2n}}}{(2n)!} \;  \frac{1}{\treepowze (k)} \, \frac{d^{2n}  \, \treepowze (k)}{d(\ln k)^{2n}} \right] \;\; ,
\end{align}
and we have parametrically $\corrbase{\zbar^{2n}} \sim \corrbase{\zbar^2}^n$. Since $\ln k' = \ln k -\zbar$, this is similar to a Taylor expansion of the power spectrum in $\ln k$ around the scale $k$. At every order in the expansion, corrections consist of two counteracting contributions. On the one hand, there is the factor $\corrbase{\zbar^2}^n$ , with
\be
  \label{eq:2_point_zbar_squared}
  \corrbase{\zbar^2} = \int\limits_{1/L}^k \frac{dq}{q} \; \treepowze (q) \quad ,
\ee
that involves a log-enhancement (even though it is suppressed by the smallness of the power spectrum). On the other hand, there are derivatives of the power spectrum that consist of slow-suppressed quantities and to which we will refer as late-time suppression. The word `late-time' indicates that derivatives of the power spectrum in eq.~\eqref{eq:2_point_series_expansion}, i.e.\ the slow-suppressed quantities, are evaluated at the scale $k$. By contrast, the quantity $\corrbase{\zbar^2}$ receives contributions from all modes in the range from $1/L$ to the scale $k$. Convergence of the series expansion depends on the ability of the late-time suppression to compensate the log-enhancement due to $\corrbase{\zbar^2}$.

We will perform an order of magnitude analysis and, hence, we do not distinguish between quantities that are of the same order in slow-roll, like the slow-roll parameters $\epsilon$ and $\eta$. Instead, we generically characterize the slow-roll suppression by an appropriate power of $\epsilon$. A derivative $d /d(\ln k)$ acting on $\treepowze$ precisely corresponds to one such power of $\epsilon$. Hence, the late-time suppression is given by
\be
 \frac{1}{\treepowze (k)} \, \frac{d^{2n}  \, \treepowze (k)}{d(\ln k)^{2n}} \sim \epsilon^{2n}(k)  \quad .
\ee
Therefore, in our order of magnitude analysis, eq.~\eqref{eq:2_point_series_expansion} can be written as
\be
  \frac{\powze (k)}{ \treepowze (k) } -1 \sim   \sum\limits_{n=1}^\infty \left( \epsilon^2(k) \corrbase{\zbar^2} \right)^n \; \; ,
\ee
and the convergence of the series expansion requires $\epsilon^2 \corrbase{\zbar^2} < 1$.

Let us first consider a weakly scale-dependent power spectrum. Here, `weakly scale-dependent' means that the power spectrum $\treepowze$ has only a very mild scale-dependence on the complete range from $1/L$ to $k$ such that eq.~\eqref{eq:2_point_zbar_squared} essentially yields
\be
  \corrbase{\zbar^2} = \treepowze \; \ln (kL) \sim \frac{H^2}{\epsilon} \; \ln (kL)  \;\; .
\ee
The logarithm is given by the number of observed e-foldings $N\simeq H t$. Therefore, it remains to verify the relation
\be
  \epsilon^2 \corrbase{\zbar^2} \sim \epsilon \; H^3 t < 1 \quad .
\ee
As shown by \cite{ArkaniHamed:2007ky, Dubovsky:2008rf}, the requirement of being in a non-eternally inflating phase constrains the time to obey $t < R \,S \sim H^{-3}$, where $R$ and $S$ are deSitter radius and entropy, respectively. Hence, the criterium for convergence reduces to $\epsilon < 1$, which is fulfilled by construction in slow-roll inflationary models. Therefore, under the assumption of a weakly scale-dependent power spectrum, the series is always converging. However, this is not surprising. The assumption of a `weakly scale-dependent' power spectrum can be made mathematically more precise by demanding that the scale-dependence of $\treepowze$ is negligible in the integral in eq.~\eqref{eq:2_point_zbar_squared}  (such that $\corrbase{\zbar^2} = \treepowze \ln(kL)$~). This yields
\be
  \label{eq:2_point_weakly_scale_dependence_condition}
  \left. \frac{1}{\treepowze (k)} \, \frac{d^n\treepowze}{d(\ln k)^n} \right|_k \; \left[ \, \ln(kL) \, \right]^n \ll 1 \qquad n>0 \quad ,
\ee
from which we could have concluded the convergence of the series in eq.~\eqref{eq:2_point_series_expansion} directly.

In spite of all that was said above, convergence breaks down for the model of chaotic inflation characterized by the power spectrum \eqref{eq:2_point_monomial_potential_tree_power_spectrum} (see also \cite{Giddings:2010nc}). The integration in the expression of $\corrbase{\zbar^2}$ in this model needs to be performed over several orders of magnitude in the scalar background field $\phi$. Therefore the slow-roll parameter $\epsilon\sim 1/\phi^2$ is changing over several orders of magnitude. This clearly violates the approximation of a weak scale-dependence. Consequently, convergence is not obvious in this model and an investigation of the convergence behavior requires a more precise evaluation of $\corrbase{\zbar^2}$. Indeed, $\corrbase{\zbar^2} \sim \int (H^2/\epsilon)\, dq/q$ is completely dominated by contributions at very early times $t_i$. Hence, the expansion parameter $\corrbase{\zbar^2}$ is much larger than $\treepowze (k) \ln(kL)$. By contrast, the coefficients in eq.~\eqref{eq:2_point_series_expansion}, i.e.\ the late-time suppression, consist of slow-roll parameters which are large compared to those at early times $t_i$. Hence, the late-time suppression cannot compensate the enhancement at early times leading to a breakdown of convergence.

In principle, this effect is also present for tensor corrections, though less severe since the power spectrum of tensor modes is not enhanced by $1/\epsilon$. Therefore, in this model the breakdown of convergence due to scalar contributions occurs first. This observation is also in agreement with the findings in \cite{Giddings:2010nc}, showing that the effect of scalars dominates.

We note that a breakdown of convergence implies that one cannot trust 
conventional perturbation theory. However, this only applies to the 
conventionally defined power spectrum at sufficiently large $L$. In our 
philosophy, one should instead consider higher-order corrections to IR-safe 
quantities like the power spectrum $\treepowze (k)$ defined in 
eq.~(\ref{eq:geomety_tree_power_spectrum}). We know that the leading-order 
corrections to this object will not be log-enhanced. While we have not 
shown this in the present paper, we expect that also higher-order 
corrections will benefit from our IR-safe definition and hence that 
conventional QFT perturbation theory, based on the smallness of $\zeta$ 
and of slow-roll parameters, will be reliable.

\section{ Three-point function and the bispectrum}

\label{sec:3_point_bispectrum}

The bispectrum accounts for the simplest contribution to non-Gaussianity. Starting from the three-point function in momentum space, one extracts the {\it bispectrum} from its connected part:
\be
  \langle\zeta_{\vec{k_1}} \zeta_{\vec{k_2}}\zeta_{\vec{k_3}} \rangle \, \equiv \, (2 \pi)^3 \deltaThreeD{\kv_1+\kv_2+\kv_3} \; B_\zeta (\kv_1, \kv_2) \qquad .
\ee
In this section, for definiteness we focus on non-Gaussianity of local form (see \cite{Wands:2010af} for a recent review)~\footnote{Other  forms of non-Gaussianity can also be described with techniques similar to the ones we are are going to develop.}. The corresponding bispectrum is well-described by
\be
  \label{locanbis}
  B_\zeta (\kv_1, \kv_2) \, = \, \frac65 \fnl(\kv_1, \kv_2) \; \left[ P_\zeta (k_1) P_\zeta (k_2) + perms \right] \quad ,
\ee
where $\fnl$ is a slowly-varying function and we have introduced the uncurly power spectrum $P_\zeta (k)=2\pi^2\,\powze (k)/k^3$. Here and henceforth, we indicate with $perms$ all non-trivial cyclic permutations of $\kv_1, \kv_2$ and $\kv_3 = -(\kv_1 + \kv_2)$~. The dependence of $P_\zeta(k)$ on the long-wavelength background modes is characterized
\be
 \label{noncps}
 P_\zeta(k)\,= \, \left\langle e^{-3\zbar} \; P_\zeta^{(0)} \left( k' \right) \right\rangle  \quad .
\ee
With the formalism of sec.~\ref{sec:geometry_of_the_reheating_surface} for higher correlation functions, we may immediately write down $\fnl$ including long-wavelength corrections:
\begin{align}
 \fnl &= \frac{5}{6} \frac{B_\zeta (\kv_1, \kv_2)}{\left[ P_\zeta (k_1) P_\zeta (k_2) + perms \right]}  \nonumber \\
      &= \frac{5}{6} \frac{ \corrbase{ \; e^{-6\zbar} \; B^{(0)}_\zeta (\kv_1', \kv_2') \; } }{ \corrbase{ \; e^{-3\zbar} \; P_\zeta^{(0)} (k_1')} \corrbase{ \; e^{-3\zbar} \; P_\zeta^{(0)} (k_2')} + perms } \quad .
      \label{eq:bispectrum_fnl_expansion}
\end{align}
The remaining task is to evaluate \eqref{eq:bispectrum_fnl_expansion}. This requires knowledge on the tree-level bispectrum $B^{(0)}_\zeta$, which is model dependent.

As an illustrative example, we consider the form 
\be
  \label{eq:bispectrum_tree_bispectrum}
  B^{(0)}_\zeta = \frac{6}{5} \left[ P_\zeta^{(0)} (k_1) P_\zeta^{(0)} (k_2) \; f_\zeta(k_3) + perms  \right] \qquad .
\ee
This tree-level bispectrum is motivated by a curvature perturbation which is given by a Gaussian part, $\zeta^G$, plus $f_\zeta (k)$ times the Gaussian part squared, i.e.\ 
\be
  \label{eq:bispectrum_gaussian_plus_gaussian_squared}
  \zeta_{\kv} = \zeta_{\kv}^{G} + f_\zeta (k) \left( \zeta^G \star \zeta^G \right)_{\kv} \quad .
\ee
Here, the operator $\star$ denotes a convolution. In concrete examples, $f_\zeta$ depends on the  scales $k$ only by means of the dependence on times of horizon exit for each mode \cite{Byrnes:2010ft, Byrnes:2009pe}. Note that the tree-level bispectrum \eqref{eq:bispectrum_tree_bispectrum} has a slightly different scale-dependence than eq.~\eqref{locanbis}. They only match for the popular assumption of $f_\zeta$ being scale-invariant or in the squeezed limit, where one scale is much smaller than the others (say $k_1 \ll k_2 , k_3$). Indeed, one has $\fnl^{(0)} = f_\zeta$ in these cases.

We stress that the bispectrum \eqref{eq:bispectrum_tree_bispectrum} neglects the presence of intrinsic non-Gaussianity in the second order scalar field fluctuations. To include this contribution, one has to apply the bispectrum given by Maldacena \cite{Maldacena:2002vr}. In order to keep equations simple, we will neglect this presence of intrinsic non-Gaussianity and apply eq.~\eqref{eq:bispectrum_tree_bispectrum}.

Note that the primary field of validity of eq.~\eqref{eq:bispectrum_tree_bispectrum} is in multi-field, e.g.\ curvaton-type, models with observable non-Gaussianity. In addition, it arises in the squeezed limit of single field slow-roll inflation. In that case, our modified $\delta N$-formalism, presented in section \ref{sec:alternative_way}, reproduces the correct result for $\fnl^{(0)}$, i.e.\ $\fnl^{(0)}= 5/12\; (1-n_\zeta)$ (see appendix \ref{AppA}). This agreement shows that, contrary to the conventional $\delta N$-formalism, our modified version of $\delta N$ provides correct results also for the 3-point function in the squeezed limit. Consequently, the following calculation is correct in the squeezed limit, even though we made the simplifying assumption of negligible intrinsic non-Gaussianity.

Proceeding as we did for the power spectrum, we perform a slow-roll expansion for the quantities inside the averages in eq.~\eqref{eq:bispectrum_fnl_expansion}, focussing on the non-vanishing contributions at leading order in slow-roll. After some calculation, this yields
\begin{align}
 \label{eq:bispectrum_fnl_result}
 \fnl &= \fnl^{(0)} \left[ 1 + \frac{\Omega(\kv_1, \kv_2, \kv_3) \; P^{(0)}_\zeta (k_1) \, P^{(0)}_\zeta (k_2)  \, f_\zeta(k_3) + perms }{P^{(0)}_\zeta (k_1) \, P^{(0)}_\zeta (k_2) f_\zeta(k_3) + perms } \right]  \\
   & = \fnl^{(0)} \left[ 1 + \frac{\Omega(\kv_1, \kv_2, \kv_3) \; k_3^3 + \Omega(\kv_3, \kv_1, \kv_2) \; k_2^3 + \Omega(\kv_2, \kv_3, \kv_1) \; k_1^3 }{k_1^3 + k_2^3 + k_3^3} \right]
   \label{eq:bispectrum_fnl_result_part2}  \\
 \fnl^{(0)} & = \frac{5}{6} \frac{B_\zeta^{(0)}(\kv_1, \kv_2) }{P_\zeta^{(0)}(k_1) P_\zeta^{(0)}(k_2) + perms }
   \label{eq:bispectrum_tree_fnl} \\[.5ex]
 \Omega(\kv_1, \kv_2, \kv_3) &= \frac{3}{20} \corrbase{\tr\gbar^2} \left[ 3 (\hat{k}_1 \cdot \hat{k}_2)^2 -1 \right] \nonumber \\
   & \phantom{=} \; - \frac{1}{20} \corrbase{\tr\gbar^2} \left\lbrace 2 \left[ 3 (\hat{k}_1 \cdot \hat{k}_2)^2 - 1 \right]  \frac{1}{\treepowze} \frac{d\treepowze}{d\ln k} + 3  \left[ (\hat{k}_1 \cdot \hat{k}_3)^2 + (\hat{k}_2 \cdot \hat{k}_3)^2 -1 \right]  \frac{1}{f_\zeta} \frac{d\, f_\zeta}{d\ln k}  \right\rbrace \nonumber \\
   & \phantom{=} \; + \frac{\corrbase{\zbar^2}}{2}  \left\lbrace \frac{1}{f_\zeta} \frac{d^2 f_\zeta}{d(\ln k)^2} + 2 \left(\frac{1}{\treepowze} \frac{d\treepowze}{d\ln k} \right)^2 + 4 \frac{1}{f_\zeta \; \treepowze} \frac{d\, f_\zeta}{d\ln k} \frac{d\treepowze}{d\ln k} \right\rbrace
 \label{eq:bispectrum_fnl_corrections} \quad .
\end{align}
Here $\hat{k}_i \cdot \hat{k}_j$ corresponds to the cosine of the angle between the vectors $\kv_i$ and $\kv_j$. $\corrbase{\tr\gbar^2}$ and $\corrbase{\zbar^2}$ are defined as before. In the previous expression for $\Omega$, the scale at which we evaluate  $\treepowze$, $f_\zeta$ and their derivatives is any one of the $k_i$: the difference among these quantities evaluated at different scale is  slow-roll suppressed with respect to the contributions we are examining. As a result, a cancellation of corrections originating from the numerator and the denominator in eq. (\ref{eq:bispectrum_fnl_result}) occurs. This removes terms containing second derivatives of the power spectrum. Moreover, it is sufficient to take into account the naive scaling $P_\zeta^{(0)}(k) \sim k^{-3}$ in eq.~\eqref{eq:bispectrum_fnl_result}, leading to the simpler form in eq.~\eqref{eq:bispectrum_fnl_result_part2}. In eq.~\eqref{eq:bispectrum_tree_fnl}, we defined the leading order non-Gaussianity parameter $\fnl^{(0)}$.

In some cases, it may be useful to perform an average over directions of the vectors. However, we note that the $\delta$-function sets constraints on this averaging procedure. As an example, we focus on the particular case of squeezed configurations, i.e.\ $k_1 \ll k_2, k_3$. For these configurations, one of the permutation terms can be dropped and the $\delta$-function requires $(\hat{k}_2 \cdot \hat{k}_3)^2=1$. The pair of unit-vectors $\hat{k}_1, \hat{k}_2$ or $\hat{k}_1, \hat{k}_3$ is statistically independent. Hence, the directional averaging gives $(\hat{k}_1 \cdot \hat{k}_2)^2 = (\hat{k}_1 \cdot \hat{k}_3)^2 = 1/3$. Therefore, having performed the directional averaging, the expression for squeezed configurations reads
\begin{align}
 \fnl & = \fnl^{(0)} \; \left[ 1 + \Omega(\kv_1, \kv_2, \kv_3) \right]  \\
 \Omega(\kv_1, \kv_2, \kv_3) & = - \frac{1}{20} \corrbase{\tr\gbar^2} \frac{1}{f_\zeta} \frac{d\, f_\zeta}{d\ln k}  \nonumber \\
  &\phantom{=} + \frac{\corrbase{\zbar^2}}{2}  \left\lbrace \frac{1}{f_\zeta} \frac{d^2 f_\zeta}{d(\ln k)^2} + 2 \left(\frac{1}{\treepowze} \frac{d\treepowze}{d\ln k} \right)^2 + 4 \frac{1}{f_\zeta \; \treepowze} \frac{d\, f_\zeta}{d\ln k} \frac{d\treepowze}{d\ln k} \right\rbrace \;\;.
\end{align}
We stress that under the directional averaging the first term on the right-hand side in eq.~\eqref{eq:bispectrum_fnl_corrections} vanishes. Indeed, we will see below that, keeping the directional information, precisely this term turns out to be the leading order correction. Therefore, this example illustrates that such procedures have to be handled with care.

Neglecting tensor fluctuations, the special case of corrections to $\fnl$ in squeezed configurations was also discussed in \cite{Giddings:2010nc}. In this case, corrections are solely given by the last line in eq.~\eqref{eq:bispectrum_fnl_corrections}, which reads
\be
  \label{eq:bispectrum_GS_corrections}
  \Omega(\kv_1, \kv_2, \kv_3) = \frac{\corrbase{\zbar^2}}{2}  \left\lbrace \frac{1}{f_\zeta} \frac{d^2 f_\zeta}{d(\ln k)^2} + 2 \left(\frac{1}{\treepowze} \frac{d\treepowze}{d\ln k} \right)^2 + 4 \frac{1}{f_\zeta \; \treepowze} \frac{d\, f_\zeta}{d\ln k} \frac{d\treepowze}{d\ln k} \right\rbrace \quad .
\ee
Our result basically agrees with the findings of \cite{Giddings:2010nc}. In order to have complete agreement, one needs to take into account the runnings of $\treepowze$ and $f_\zeta$ in the calculation of the bispectrum in \cite{Giddings:2010nc} (~their eqs.\ (5.9)-(5.11)~). Including this effect, their final formula for $\fnl$, eq.\ (5.14), slightly changes. Like in our findings, the running of the power spectra from the denominator disappears since it cancels against corresponding terms from the numerator. The effect of the running of $f_\zeta$ appears precisely as the first term on the right-hand side of eq.~\eqref{eq:bispectrum_GS_corrections}. The second and third term on the right-hand side of eq.~\eqref{eq:bispectrum_GS_corrections} are already present in eq.\ (5.14) in \cite{Giddings:2010nc}.
\smallskip

We now return to the general form of corrections to $\fnl$, i.e.\ eqs.~\eqref{eq:bispectrum_fnl_result}-\eqref{eq:bispectrum_fnl_corrections}. Remarkably, we find that in single-field, slow-roll inflation tensors provide the dominant contribution in slow-roll, i.e.\ the first term on the RHS in eq.~\eqref{eq:bispectrum_fnl_corrections}.  This contribution results in a correction proportional to first order slow-roll parameters, while the others are of second order. Indeed, we observe that, at leading order in slow-roll, the dominant contribution is originating from the prefactor
\be
  \langle m(\hat k_1) m(\hat k_2)\rangle \,=\,1+\frac{3}{20}\,\langle {\rm tr} \bar \gamma^2\rangle \,\left[3 (\hat{k}_1 \cdot \hat{k}_2)^2 - 1 \right] \qquad ,
\ee
which multiplies the tree-level bispectrum. From this we find that the dominant log-enhanced contribution to $\fnl$, in a slow-roll
expansion, reads
\be
  \fnl \,= \,\fnl^{(0)} \, \left[ 1+ \frac{6r}{5} \;  \treepowze \; \ln(kL) \; \frac{ (3 (\hat{k}_1 \cdot \hat{k}_2)^2 - 1 ) \;  k_3^3 + (3 (\hat{k}_3 \cdot \hat{k}_1)^2 - 1 ) \;  k_2^3 + (3 (\hat{k}_2 \cdot \hat{k}_3)^2 - 1 ) \;  k_1^3 }{ k_1^3 + k_2^3 + k_3^3 }   \right] \;\; .
\ee
Interestingly, these log-enhanced contributions to $\fnl$ do not depend on the tilt of the power spectrum, and so are also present for spectral index equal to one.

In conclusion, log-enhanced contributions to $\fnl$ can be expressed in terms of observable  quantities. Tensor contributions are proportional to first order slow-roll parameters, and are  suppressed by the tree-level power spectrum. Very similar results hold for parameters associated to the trispectrum, $g_{\rm NL}$ and $\tau_{\rm NL}$. It is straightforward to obtain them proceeding exactly  as done in this section. 
\smallskip

Local non-Gaussianity in single field, slow-roll inflation turns out to be small. On the other hand, models, in which a second field takes part in the generation of curvature perturbations as in the curvaton scenario, can lead to large values of $\fnl$ (see e.g. \cite{Lyth:2002my}). In the approximation in which only the curvaton field is responsible 
for curvature perturbations, the tree level bispectrum reads
\be
  B^{(0)}_\zeta (\kv_1, \kv_2) = f_{\sigma}(\kv_1, \kv_2) \; \left[ \frac{2\pi^2}{k_1^3}  \mathcal{P}_\sigma (k_1) \; \frac{2\pi^2}{k_2^3} \mathcal{P}_\sigma (k_2) + perms  \right] \qquad ,
\ee
where $\sigma$ indicates the curvaton field. Our approach can be applied also to this case, although it requires additional work to calculate the contributions of long-wavelength scalar modes to inflationary observables, since more than one scalar field is present. We outline a method to do this in Appendix \ref{AppA}, but a more complete discussion of this issue is left for future work. In this case, enhancement effects associated with long-wavelength modes could turn out to be more important than the ones discussed so far.

\section{Conclusions}

\label{sec:conclusions}

We have considered IR effects associated with backreaction of long-wavelength scalar and tensor modes in inflationary backgrounds. We proposed an infrared-safe definition of correlation functions involving curvature fluctuations, with no sensitivity on long-wavelength contributions. The essential idea was to make use of the proper invariant distance on the reheating surface where the curvature perturbation is evaluated. By using the invariant distance, one automatically absorbs longer wavelength modes in the background and obtains $n$-point functions for the curvature perturbation that are free from IR contributions associated with long-wavelength modes. We showed how to re-interpret our results in terms of conventionally defined $n$-point functions. This allowed us to provide closed expressions for the latter that manifestly exhibit the dependence on long-wavelength modes. In our approach, IR corrections automatically emerge in a resummed, all-orders form. We then applied our approach to the analysis of inflationary observables built from (conventionally defined) two- and three-point functions of the curvature perturbation. We showed how to compute the leading scalar and tensor IR effects on the power spectrum and on the bispectrum, in single field, slow-roll inflation. Our corrections to the power spectrum (both from long-wavelength scalar and tensor modes) and to $\fnl$ (from long-wavelength scalar modes) agree (essentially) with Giddings and Sloth \cite{Giddings:2010nc} (obtained by somewhat different methods). The advantage of our approach is that it directly provides resummed, all-orders expressions. We extend \cite{Giddings:2010nc} by tensor corrections to $\fnl$. This is, in fact, the dominant piece! We also explicitly computed, in a specific inflationary model, the complete, all-orders expression for scalar long-wavelength contributions to inflationary observables. Furthermore, we analyzed the question of convergence of IR corrections. Using entropy bounds given in \cite{ArkaniHamed:2007ky,Dubovsky:2008rf}, we found that for a weak scale-dependence the convergence of the series of IR corrections is guaranteed. However, despite the existence of these entropy bounds and the fulfillment of slow-roll conditions, the convergence of the IR-correction series may break down if the scale-dependence is not sufficiently weak.

Summarizing, we have provided a simple formalism to calculate and investigate inflationary IR corrections. Maybe more importantly, we have provided simple definitions of IR-safe correlation functions which make it possible to avoid IR enhancement altogether.

We have also shown that in all cases, where the $\delta N$-formalism is applicable, our results can be equivalently obtained in terms of a suitable generalization of the $\delta N$-formalism, extending the discussion of \cite{Byrnes:2010yc}. In the present work, we included the effects of graviton long-wavelength modes, and we explained how to calculate
IR contributions to arbitrary $n$-point functions involving curvature perturbations.  

A natural question is how to extend our results to the case in which more than one field plays an active role in generating the  curvature perturbations. In this case, IR
effects might play a role more important than the one for single field inflation. We outlined in an Appendix a method to treat this problem, but we leave a more complete discussion for future work.

\section*{Acknowledgments}

We thank Chris Byrnes for many useful conversations and helpful discussions. A.H.\ acknowledges the hospitality of the Perimeter Institute during the workshop ``IR Issues and Loops in de Sitter Space'' and related discussions with Steven Giddings, Martin Sloth, Takahiro Tanaka and Yuko Urakawa. This work was supported by the German Research Foundation (DFG) within the Transregional Collaborative Research Centre TR33 ``The Dark Universe''. G.T.\ is founded by an STFC Advanced Fellowship (Ref Number ST/H005498/1). M.G.\ acknowledges support from the \textit{Studienstiftung des Deutschen Volkes}.

\begin{appendix}

\section{Comparison of $\powze$ and $\treepowze$}
\label{app:calculation_IR_safe_power_spectrum}

The definitions of the IR-sensitive power spectrum $\powze$ and the IR-safe power spectrum $\treepowze$ are
\begin{align}
 \powze (k)  & =  \frac{k^3}{2\pi^2} \int\limits d^3 y \; e^{-i\kv\yv} \, \corr{\zeta(\xv)}{\zeta(\xv + \yv)}  \\
 \treepowze (k) &= \frac{k^3}{2\pi^2} \int\limits d^3 z \; e^{-i\kv\zv} \, \corr{ \zeta(\xv)}{\zeta(\xv+ e^{-\zbar} \, e^{-\gbar/2} \, \zv)} \qquad .
\end{align}
Here, $\zv$ and $\yv$ are related by $z^i = e^{\zbar} \left( e^{\gbar/2} \right)_{ij} \, y^j$. Comparing the two yields
\begin{align}
 \powze (k) &= \left\langle \; \frac{k^3}{2\pi^2} \int\limits d^3 y \; e^{-i\kv\yv} \, \zeta(\xv) \; \zeta(\xv + \yv) \; \right\rangle \\
  &= \left\langle \; \frac{k^3}{2\pi^2} \int\limits d^3 y \; e^{-i\kv\yv} \, \zeta(\xv) \; \zeta(\xv + e^{-\zbar} e^{-\gbar/2}(e^{\zbar} e^{\gbar/2}\yv) \,) \; \right\rangle  \\
            &= \left\langle \; \frac{ k^3 }{2\pi^2} \; e^{-3\zbar}  \int d^3z \; \exp\{-i (e^{-\zbar} e^{-\gbar/2} \kv) \zv \}  \;\; \zeta(\xv) \; \zeta(\xv + e^{-\zbar} e^{-\gbar/2}\zv \,) \; \right\rangle  \\
            &= \left\langle \;  \left[  \left(e^{-\gbar}\right)_{ij} \hat{k}_i \hat{k}_j \right]^{-3/2}  \; \treepowze \left( e^{-\zbar} e^{-\gbar/2} \kv \,\right)  \right\rangle  \qquad .
\end{align}
In the first line, we included the integral and prefactors in the average. Note that this does not affect the averaging process over pairs of points separated by the coordinate vector $\yv$. From the second to the third line, we performed a coordinate transformation of the integration variable from $y$ to $z$. Since the determinant of $e^{\gbar}$ is one, tensor fluctuations do not effect this transformation. Therefore, only scalar fluctuations appear as a prefactor in the third line. Consequently, we need to add tensor fluctuations by hand in this prefactor, in order to express the third line in terms of the IR-safe power spectrum $\treepowze$. This results in the prefactor in the last line, which only consists of tensor fluctuations. In this last line, the vector $\hat{k}$ is a unit vector in $\kv$-direction and the average is performed over the background quantities $\zbar(\xv)$ and $\gbar_{ij}(\xv)$.

\section{Extension of $\delta N$-formalism}\label{AppA}

In this appendix, we discuss in more detail how our results can be understood in terms of a $\delta N$ approach. In a previous paper \cite{Byrnes:2010yc}, written in collaboration with Byrnes and Nurmi, we showed how a suitable extension of the $\delta N$-formalism allows for the computation of leading-log contributions to the power spectrum, due to scalar long-wavelength fluctuations. Here, we extend our work to include tensor modes and to compute log-enhanced corrections to non-Gaussianity parameters. 

Let us start with single field inflation. For this purpose, we will adopt the same gauge as in section \ref{sec:alternative_way}. By means of the $\delta N$-formalism, the curvature perturbation $\zeta$ can then be expressed in terms of the number of e-foldings evaluated on a background given by the scalar field $\phi$ and its perturbation $\delta\phi$:
\be
  \zeta\,=\,N\left[\phi+\delta \phi\right]-\langle N \rangle \quad .
\ee
The previous expression admits an expansion in terms of scalar fluctuations
\be
 \label{anssbag}
 \zeta_{\kv}\,=\,N_\phi(k') \delta_{\kv} \phi(k', 
 \bar g_{ab})+\frac12 N_{\phi\phi}(k')
 \left[
  \left(
  \delta \phi\star
  \delta \phi\right)_{\kv} (k', \bar g_{ab})-\langle
  \delta \phi\star
 \delta \phi
 \rangle
 \right]+... \quad ,
\ee
where we use the notation of the main text, and the Gaussian, first-order scalar fluctuation $\delta_{\vec k} \phi$ is given in eq.~\eqref{newflu}. The effect of long-wavelength modes is encoded in the shift of the time of horizon exit, and in the function $m$, contained in the expression of $\delta_{\vec k} \phi$. Long-wavelength mode contributions are controlled by the averaged quantities $\zbar$ and $\gbar_{ij}$.

With the previous expression, we neglect intrinsic non-Gaussianity of $\delta_{\vec k} \phi$. Hence, this formalism is only applicable in situations where this is negligible. However, this condition is fulfilled in several models, e.g.\ models in which non-Gaussianity of the local form can acquire sizeable values as in multiple field inflation or curvaton-like mechanisms \cite{Suyama:2010uj}. In light of these models, it is worthwhile to develop formalisms that neglect the presence of second order fluctuations.

Using the previous formula, it is straightforward to compute $n$-point functions of curvature perturbations, and compute leading log-enhanced corrections to inflationary observables. It is important to stress that, due to the choice of a gauge with $\delta_{\qv} \phi = 0$ for $q \ll k$, convolutions appearing in the second term of eq.~\eqref{anssbag} do not involve integration over all the modes, but have a lower cut-off slightly below the scale $k$. This implies that convolutions, when appearing in $n$-point functions, do not provide further log-enhanced contributions with respect to the ones associated with long-wavelength background modes. All the IR dependence is then contained in the quantities $\zbar$ and $\gbar$. 
\bigskip

As an example, let us work out explicitly the expression for the three-point function in the squeezed limit, including the effects of long-wavelength modes, using eq.~\eqref{anssbag}. Our method is similar to \cite{Allen:2005ye}. The second term on the right-hand side in eq.~\eqref{anssbag} is irrelevant for squeezed configurations and will be neglected. The contribution to the bispectrum is 
\be
  \corrbase{ \zeta_{\kv_1} \zeta_{\kv_2} \zeta_{\kv_3} } = \frac{1}{\sqrt{8 k_1^3 k_2^3 k_3^3 }} \;\;
    \corrbase{ \left[ m^{\frac{1}{2}}(\hat{k}_1) \, (\None H)(k_1 ') \; a_{\kv_1} \right] 
               \left[ m^{\frac{1}{2}}(\hat{k}_2) \, (\None H)(k_2 ') \; a_{\kv_2} \right] 
               \left[ m^{\frac{1}{2}}(\hat{k}_3) \, (\None H)(k_3 ') \; a_{\kv_3} \right]   }  \;\; .
\ee
In the limit in which $k_1\ll k_2 \simeq k_3$, the size of the vector $k_1$ is comparable to the size of the long-wavelength modes relative to the vectors $k_2$ and $k_3$. The latter are included in the shift of the time of horizon exit $t_{k_2'}$  and on $m(\hat k_2)$, respectively, $t_{k_3'}$  and $m(\hat k_3)$. Taking into account this fact, and using Wick's theorem, we can write in this limit the following non-vanishing contribution
\be
  \label{eq:AppA_3_point_correlator_squeezed}
  \corrbase{ \zeta_{\kv_1} \zeta_{\kv_2} \zeta_{\kv_3} } = (2\pi)^3 \deltaThreeD{\kv_2 + \kv_3 } \; \frac{1}{2 k_2^3 \sqrt{2 k_1^3}} \;
     \corrbase{ \left[ m^{\frac{1}{2}}(\hat{k}_1) (\None H)(k_1 ') \; a_{\kv_1} \right] \left[ m(\hat{k}_2)  \left( \None H \right)^2(k_2 ') \right] }  \;\; .
\ee
Since $k_1 \ll k_2$, the only possibility to contract $a_{\kv_1}$ is the background contribution originating from $(\None H)^2(k_2')$. By expanding the latter and by means of the definition of $\zbar$, this contraction yields
\be
  \corr{ a_{\kv_1} }{ (\None H)^2(k_2') } = - \corr{a_{\kv_1}}{ \zbar} \; \left. \frac{d \left(\None H \right)^2}{d\ln k} \right|_{k_2 '} = - m^{\frac{1}{2}}(\hat{k}_1) \, \frac{(\None H)(k_1 ')}{\sqrt{2k_1^3}}  \;\; \left. \frac{d \left(\None H \right)^2}{d\ln k} \right|_{k_2 '}  \quad .
\ee
Therefore, we find for the bispectrum:
\be
  \label{eq:AppA_bispectrum}
  B_\zeta (\kv_1, \kv_2) = \frac{-1}{4(k_1 k_2)^3} \; \corrbase{ \left[m(\hat{k}_1) (\None H)^2 (k_1 ') \right] \left[ m(\hat{k}_2) (\None H)^2 (k_2 ') \right] \left. \frac{d \ln\treepowze}{d\ln k} \right|_{k_2 '}   } \;\; .
\ee
At leading order, we neglect the contribution from the background ($k_i ' \rightarrow k_i$ and $m(\hat{k}_i)=1$). This provides the following tree-level result for the non-Gaussianity parameter
\be
 \label{eq:fnl_squeezed_limit_LO}
 \fnl^{(0)} (k) = -\frac{5}{12} \left. \frac{d \ln\treepowze}{d\ln k} \right|_{k} = \frac{5}{12} (1-n_\zeta (k) ) \quad ,
\ee
that is Maldacena's consistency relation \cite{Maldacena:2002vr, Creminelli:2004yq}. Hence, the complete form for $\fnl$, obtained from formula \eqref{eq:AppA_bispectrum}, can be expressed through $\fnl^{(0)}$, giving
\be
  \fnl = \frac{ \corrbase{ m(\hat{k}_1) \treepowze (k_1 ') \;\;  m(\hat{k}_2) \treepowze (k_2 ') \;\; \fnl^{(0)} (k_2 ') } }{ \corrbase{ m(\hat{k}_1) \treepowze (k_1 ') } \; \corrbase{ m(\hat{k}_2) \treepowze (k_2 ') } }  \;\; ,
\ee
in agreement with formula \eqref{eq:bispectrum_fnl_expansion} in the squeezed limit.  We can then proceed as done in the main text to extract leading log-enhanced contributions.

\bigskip
Let us briefly discuss the case in which multiple scalar fields affect the curvature perturbation. The $\delta N$-formalism is very well suited to study this case, as discussed in the original paper by Sasaki and Stewart \cite{Sasaki:1995aw}. We adopt a gauge with vanishing scalar metric fluctuations, i.e.\ the long-wavelength modes of the scalar field are not vanishing. The curvature perturbation (that in the case of multiple fields is not generally conserved) can be expressed as an expansion in terms of all the scalar fields involved
\be
 \label{ansamulti}
 \zeta_{\kv}(t_f)\,=\,N_I\left[t_f, \{\phi_0\} \right] \, \delta_{\kv} \phi^I\left[ \{\phi_0\} \right] + \frac12 N_{IJ}\left[t_f, \{\phi_0\} \right] \, \left\{\left(\delta \phi^I \star \delta \phi^J\right)_{\kv} \left[ \{\phi_0\} \right] - \langle \delta \phi^I \star \delta \phi^J\rangle\right\}+\cdots  \;\; ,
\ee
where the capital latin indices of $N$ denote derivatives w.r.t.\ the scalar fields and summation over repeated indices is understood. Here, $\{\phi_0\}$ denotes the dependence on the homogeneous values $\phi_0^I$ of the scalar fields. As in Sasaki and Stewart, we have replaced the dependence on the time of horizon exit $t_k$ of the various functions, with the value of homogeneous solutions of the scalar equations at $t_k$: $\phi_0^I\equiv\phi_0^I(t_k)$~.

Then, the inclusion of the effects of scalar and tensor long-wavelength modes can be done as in the previous sections, although the procedure is a bit more laborious. We express a given function $\phi^I(t,\vec{x})$, the solution of the field equations, as
\be
 \phi^I(t,\vec{x})\,=\,\phi^I_0(t)+ \delta \bar \phi^I(t) + \delta \phi^I(t,\vec{x})  \quad ,
\ee
where $\delta \bar \phi^I(t)$ is an average over long-wavelength modes, similar to the ones we performed in the main text. The effect  of long-wavelength scalar fluctuations is to shift the values of $\phi_0^I$, that appear in eq.~\eqref{ansamulti}, to $\phi_0^I+\delta \bar \phi^I$. In a sense, they play the same role of shifting the time of horizon exit, although with multiple fields there is not a one to one correspondence between time and values of the scalar solution. After passing to momentum space, the inclusion of long-wavelength scalar perturbations implies that the expansion in eq.~\eqref{ansamulti} becomes
\begin{align}
 \zeta_{\kv}(t_f) &=  N_I\left[t_f, \{\phi_0 + \delta \bar \phi \}  \right]\, \delta_{\kv} \phi^I\left[ \{\phi_0 + \delta \bar \phi \} \right]\nonumber\\
      &\,\, + \frac12 N_{IJ}\left[t_f, \{\phi_0 + \delta \bar \phi \} \right]\,\left\{\left(\delta \phi^I  \star \delta\phi^J\right)_{\kv}\left[ \{\phi_0 + \delta \bar \phi \} \right] - \langle \delta \phi^I \star \delta \phi^J\rangle\right\} + \cdots
 \label{ansamultilw}
\end{align}
to take into account the shifts of the homogeneous solution of the scalar fields. The inclusion of tensor long-wavelength contribution, on the other hand, is very simple: since correlations between tensor and scalar modes vanish, the effect of tensors is precisely identical to that discussed in the previous sections. It can be taken into account with a proper redefinition of the time of horizon exit of a given mode $t_k\to t_{k'}$.  One can then repeat in this case the very same steps that we took in the previous sections, generalizing our results to multiple fields. This will be done in future work, where we will also discuss in this context the possibility of having large non-Gaussianity from loop effects \cite{Boubekeur:2005fj}, with sizeable  scale-dependence \cite{Suyama:2008nt, KLR}.

\end{appendix}

\end{document}